\documentclass[a4paper,11pt]{article}
\pdfoutput=1 

\usepackage{jheppub} 

\usepackage[T1]{fontenc} 
\usepackage
 {
   amsmath,amssymb,graphicx,tabularx
 }

\usepackage{multirow}
\usepackage{caption,subcaption} 
\usepackage{booktabs}
\usepackage{url}
\usepackage{hyperref}

\def\rmd{\mathrm{d}}
\def\lag{\mathcal{L}}

\def\lq{\left[} 
\def\rq{\right]} 
\def\rg{\right\}} 
\def\lg{\left\{} 
\def\({\left(} 
\def\){\right)} 
\def\realpt{\textrm{Re}\,}

\def\mat{\mathcal{M}}
\def\half{\frac{1}{2}}

\usepackage{color}

\title{\boldmath Higgs decay into four charged leptons in the presence of dimension--six operators}

\author[a,b]{Stefano Boselli}
\author[a]{Carlo M. Carloni Calame}
\author[c,a]{Guido Montagna}
\author[a]{Oreste Nicrosini}
\author[a]{Fulvio Piccinini}
\author[a,d]{and Ambresh Shivaji}

\affiliation[a]{INFN, Sezione di Pavia, Via A. Bassi 6, 27100 Pavia, Italy}
\affiliation[b]{Cavendish Laboratory, University of Cambridge, Cambridge CB3 0HE, UK}
\affiliation[c]{Dipartimento di Fisica, Universit\`a di Pavia, and INFN, Sezione di Pavia, 
Via A. Bassi 6, 27100 Pavia, Italy}
\affiliation[d] {Centre for Cosmology, Particle Physics and Phenomenology (CP3), 
               \\ Universit\'{e} Catholique de Louvain, B-1348 Louvain-la-Neuve, Belgium}

\emailAdd{sb2253@cam.ac.uk}
\emailAdd{carlo.carloni.calame@pv.infn.it}
\emailAdd{guido.montagna@pv.infn.it}
\emailAdd{oreste.nicrosini@pv.infn.it}
\emailAdd{fulvio.piccinini@pv.infn.it}
\emailAdd{ambresh.shivaji@pv.infn.it}

\preprint{CP3-17-07}

\abstract{
We study the indirect effects of New Physics 
in the Higgs decay into four charged leptons,
using an Effective 
Field Theory (EFT) approach to Higgs  
interactions. We evaluate the deviations induced by the EFT
dimension--six operators 
in observables like partial decay width and various kinematic distributions, 
including angular observables, and compare them
with the contribution of the full SM electroweak corrections. 
The calculation is implemented in an improved version of the event
generator {\tt Hto4l}, which 
can provide predictions in terms of different EFT-bases and 
is available for data analysis at the LHC. 
We also perform a phenomenological study in order to assess the benefits coming from the inclusion of differential information in the future analyses of very precise data which will be collected during the high luminosity phase of the LHC.}


\begin{document}
\maketitle
\flushbottom


\section{Introduction}
Now that a scalar particle, resembling the Higgs boson of the Standard
Model (SM), has been discovered~\cite{Aad:2012tfa,
  Chatrchyan:2012ufa}, the characterization of its
properties represents one of the major tasks of the LHC  
physics programme. Besides the intrinsic importance of confirming that
the new $125\,\mathrm{GeV}$ spin-0 boson is the Higgs particle of the
SM, the precise measurement of its
properties represents the opportunity to search for indirect hints of
new physics (NP). Up to now experimental analyses have
  extracted bounds on NP parameters in the  
  so called $\kappa-$ framework, which considers modifications
  proportional to SM
  couplings~\cite{LHCHiggsCrossSectionWorkingGroup:2012nn,
    Heinemeyer:2013tqa}. However, the $\kappa-$  
  framework does not provide a gauge invariant parametrization of NP and it cannot
  capture the effects of physics beyond SM (BSM) on kinematic
  distributions. The current  
  experimental bounds allow a deviation of 10\% in the Higgs to gauge
  bosons ($HVV$) and about 
  20\% deviation in Higgs fermion couplings. 

Given the lack of a clear evidence of NP signals 
in the LHC data already analyzed,  
it is reasonable to assume that the scale $\Lambda$, where new
particles would eventually appear, is well separated from the energy scale
of the SM spectrum. If this is the case, physics at the electroweak
(EW) scale can be adequately described by Effective Field Theory (EFT)
methods. The building blocks of the EFT Lagrangian are the SM fields.
The low-energy effects of new possible heavy degrees of freedom are
captured by effective operators with mass dimension $D$ larger than
four. Since it provides a  
model-independent parametrization of possible NP effects, the EFT
approach has become the phenomenological standard for the study of
indirect signals of NP. Regarding the Higgs sector, the majority of
these studies have interpreted the LHC data on Higgs production and decay modes
to derive constraints on the $D = 6$ parameters. It should be noted that 
these constraints do depend on certain model dependent assumptions.
Model independent approaches to Higgs physics 
have been also applied to differential cross sections in order to
investigate their resolving power to extract 
information on the presence of anomalous couplings. In particular, due to its 
particularly clean signature and non-trivial kinematics, the Higgs decay into
four leptons, i.e. $H \to Z Z^* \to 4 \ell$, has been considered in a number 
of works appeared in the literature~\cite{Stolarski:2012ps,Chen:2012jy,Chen:2013ejz, Beneke:2014sba, Gonzalez-Alonso:2014eva,
  Chen:2014pia,Gonzalez-Alonso:2015bha,Bordone:2015nqa}.
The signal strength in $H \to ZZ^*$ and in the gluon-gluon fusion (ggF) production 
channel after combining the CMS and ATLAS results is $\mu =
1.13^{+0.34}_{-0.31}$~\cite{Khachatryan:2016vau}. 

In Refs.~\cite{Stolarski:2012ps,Chen:2012jy,Chen:2013ejz}  
NP effects in $H \to 4\ell$ decays are parametrized in terms of
specific anomalous Higgs vertices, while in
Refs.~\cite{Gonzalez-Alonso:2014eva,Gonzalez-Alonso:2015bha,Bordone:2015nqa}
the language of pseudo-observables (PO) is adopted. Finally, in 
Ref.~\cite{Beneke:2014sba}, the observability of anomalous Higgs couplings in the 
$H \to Z(\to \ell^+\ell^-) \ell^+\ell^-$ channel has been studied in a EFT framework 
by considering the differential decay width $d\Gamma/dq^2$, as well as the relevant angular asymmetries. 

While in EFT new gauge-invariant operators are added to the SM Lagrangian,
the POs provide a parametrization of NP effects at the 
amplitude level and consequently they are process specific. On the
other hand, the PO approach is more general, in the sense that it does
not require any assumption on the underlying UV-complete theory. It is
important to stress that the connections among POs of different
observables become transparent once the mapping from 
EFT Wilson coefficients to POs has been set up.

In this paper, we study the $H \to 4\ell$ decay in the Standard Model
Effective Field Theory (SMEFT) framework. 
In particular, we perform a reanalysis on the effects of the
effective operators entering $H \to 4\ell$ decay channel both at the
level of integrated partial width and on the relevant and
experimentally accessible distributions. We compare the NP effects
with the contributions of the full SM electroweak corrections.
We also perform a phenomenological study 
in view of the outstanding integrated luminosity which is expected 
to be reached with the High Luminosity LHC (HL-LHC) project ($3\,\mathrm{ab}^{-1}$), that 
will allow to test the SM validity at a precision 
level which has never been achieved before. With this study we aim 
at highlighting the importance of angular distributions in constraining 
$D=6$ Wilson coefficients.

The rest of the paper is organized as follows. In Section 2, we 
introduce the phenomenological EFT Lagrangian in the so-called Higgs
basis~\cite{deFlorian:2016spz}, 
which is advocated in the literature to study the NP signatures in the
Higgs sector. In Section 3, we provide information on the 
$H \to 4\ell$ matrix elements implemented in the new version of the
{\tt Hto4l}\footnote{The code can be downloaded at the web page:
  \url{http://www.pv.infn.it/hepcomplex/hto4l.html}}
code~\cite{Boselli:2015aha}.    
Our numerical results and our study in the context of HL-LHC 
are presented in Section 4. We draw our conclusions in Section 5.
Further details are provided in the Appendices. In
Appendix~\ref{appen:H4e}, we present our results for the $H \to 4e$
integrated partial width. In Appendix~\ref{appen:dictionary} we
detail the computation of the  $H \to 4\ell$ BSM matrix elements in
the Warsaw basis~\cite{Grzadkowski:2010es} and the SILH
basis~\cite{Giudice:2007fh, Contino:2013kra}. In Appendix~\ref{appen:signal-strengths} we collect the formulae used in the analysis outlined in Section 4.

\section{Theoretical framework}

As mentioned above, the EFT approach is based on the hypothesis that
the scale $\Lambda$ of NP is much heavier than the EW scale. 
In this framework  the decoupling of new particles is
described by the Appelquist-Carazzone
theorem~\cite{Appelquist:1974tg}.
Once the heavy
degrees of freedom have been integrated out, the low-energy
effects of new particles are captured by an arbitrary number of
effective operators. The resulting effective Lagrangian takes the form
\begin{equation}
  \lag_{\mathrm{EFT}} = \lag_{\mathrm{SM}} +
  \frac{1}{\Lambda}\sum_i c_i^{(5)} {\cal O}_i^{(5)} +
  \frac{1}{\Lambda^2}\sum_i c_i^{(6)} {\cal O}_i^{(6)} + \cdots, 
  \label{eqn:eftlag}
\end{equation}
where $\lag_\mathrm{SM}$ in
Eq.~(\ref{eqn:eftlag}) is the
SM Lagrangian and it represents the lowest-order term of a 
series in the canonical dimension $D$. Each consecutive
term is suppressed by larger powers of the NP scale $\Lambda$.  
The main features of the SM Lagrangian still hold in the EFT
expansion. The field content is the same of the SM. 
Each higher-dimensional operator is invariant under the local
$SU(3)_C \otimes SU(2)_L \otimes U(1)_Y$
symmetry. Moreover, in linear EFT
 the spontaneous breaking of the
$SU(2)_L \otimes U(1)_Y$ down to $U(1)_{em}$ 
arises from the non-vanishing vacuum expectation
value (vev) of the complex Higgs doublet. Neglecting the $D=5$
lepton flavor violating operator~\cite{PhysRevLett.43.1566}, the
leading BSM effects are expected 
to be captured by $D=6$ operators.

Under the hypotheses of
lepton and baryon number conservation, flavor universality 
and a linear realization of the EWSB, all possible BSM deviations 
can be parametrized by a basis of 59 $D=6$ CP-even
operators and 6 
additional CP-odd bosonic operators.  
Different bases of $D=6$ operators, which are related by 
equations-of-motion for fields, 
have been proposed in the 
literature. 
The most popular choices are the \textit{Warsaw 
basis}~\cite{Grzadkowski:2010es} and
the \textit{SILH (Strong Interacting Light Higgs) 
basis}~\cite{Giudice:2007fh, Contino:2013kra}. 
The choice of the basis is usually led by the convenience to
minimize the number of operators that are necessary 
to parametrize the BSM effects on a given class of processes.
However, since the operators of these two bases are manifestly 
invariant under the $SU(2)_L \otimes U(1)_Y$ symmetry, the
connection between Wilson coefficients and phenomenology 
can be rather cumbersome.  
In this work we adopt the so-called 
\textit{Higgs basis}~\cite{deFlorian:2016spz}, which has been designed
to parametrize the effects  of new physics in the Higgs sector in a more 
transparent way. As in the \textit{BSM primaries}~\cite{Gupta:2014rxa}, 
the Higgs basis operators are written in terms of mass eigenstates.
It has been argued that the coefficients of this parametrization of
NP can be obtained as a linear transformation
from any other basis. These transformations 
are chosen in order to map particular combinations of Wilson
coefficients of a given basis
into a subset of anomalous couplings of the mass-eigenstates
Lagrangian extended to $D=6$ operators. 
These are called \textit{independent couplings}. 
The number of independent couplings is the same of any other basis. 
Once a maximal subset of independent couplings has been
identified, the remaining \textit{dependent couplings} can be written
as linear combinations of the independent ones. 
We would like to point out that the Higgs basis is advocated 
in the literature to perform the leading order EFT analyses of the Higgs 
data. A complete picture of next-to-leading order EFT calculations in the 
Higgs basis is not yet clear~\cite{deFlorian:2016spz}.

In this section we limit 
ourselves to describe the parts of the effective Lagrangian 
which are relevant for the Higgs decay into four leptons. 
For the derivation of the complete effective Lagrangian in the
  Higgs basis framework we refer to~\cite{deFlorian:2016spz}.
All the kinetic terms are canonically
normalized and there  is no $Z$-$\gamma$ kinetic mixing. The kinetic
and mass terms for the 
gauge bosons are the same of the SM, except the $W$ boson mass, 
which receives a correction of the form:
\begin{equation}
\lag^\mathrm{kinetic}_{D=6} = 2\delta m \frac{g_2^2 v^2}{4} W^+_\mu W^{-\mu}.
\label{eqn:deltam}
\end{equation}
Although the precision measurement of $W$ mass gives the possibility to
derive information on BSM physics and an EFT framework can be used in
this context~\cite{Bjorn:2016zlr}, it is important to stress that $\delta m$ is
presently very well constrained by experiments:
$\delta m = (2.6 \pm 1.9) \cdot 10^{-4}$  
\cite{Efrati:2015eaa}, so that the effects proportional
to $\delta m$ would be irrelevant for Higgs physics. Moreover, if the
underlying UV-complete theory preserves the custodial symmetry,
$\delta m = 0$ by hypothesis. For these reasons $\delta m = 0$ is
assumed in the following analysis.

The operators giving rise to anomalous contributions entering 
the Higgs decay into four charged leptons can be divided in five
classes. The first and most relevant class includes the effective
operators affecting the Higgs couplings to gauge bosons. Regarding the
neutral sector, the effective Lagrangian takes the form: 

\begin{equation}
\begin{split}
&\lag^{HVV}_{D=6} = \frac{H}{v} \lq \vphantom{\half} \right. (1+\delta c_Z) \frac{1}{4} (g_1^2+g_2^2)v^2 Z_\mu Z^\mu + \\
&+ c_{\gamma\gamma} \frac{e^2}{4} A_{\mu\nu}A^{\mu\nu} 
+ c_{Z\gamma} \frac{e\sqrt{g_1^2 + g_2^2}}{2} Z_{\mu\nu}A^{\mu\nu} 
+ c_{ZZ} \frac{g_1^2 + g_2^2}{4} Z_{\mu\nu}Z^{\mu\nu} + \\
&+ c_{Z\square} g_2^2 Z_\mu \partial_\nu Z^{\mu\nu} +
c_{\gamma\square} g_1 g_2  Z_\mu \partial_\nu A^{\mu\nu} + \\ 
& + \tilde{c}_{\gamma\gamma} \frac{e^2}{4} A_{\mu\nu}\tilde{A}^{\mu\nu}  
+ \tilde{c}_{Z\gamma} \frac{e\sqrt{g_1^2 + g_2^2}}{2} Z_{\mu\nu}\tilde{A}^{\mu\nu} 
+ \tilde{c}_{ZZ} \frac{g_1^2 + g_2^2}{4} Z_{\mu\nu}\tilde{Z}^{\mu\nu}
\left. \vphantom{\half} \rq,
\label{eqn:Hvv}
\end{split}
\end{equation}
where the convention to absorb the suppression factor
$1/\Lambda^2$ in the effective coefficients has been adopted. 
In the above, $V_{\mu\nu} = \partial_\mu V_\nu - \partial_\nu V_\mu$ and, 
${\tilde V}_{\mu\nu} = \frac{1}{2}\epsilon_{\mu\nu\rho\sigma}V^{\rho\sigma}$ 
for both $V=A,Z$.
 $g_1$ and $g_2$ are coupling parameters of the $U(1)_Y$ and $SU(2)_L$ gauge
 groups, respectively.
Of the six CP-even couplings in Eq.~(\ref{eqn:Hvv}) only five are
independent. We choose $c_{\gamma\square}$ as dependent coupling,
which is then expressed as the following linear combination: 
\begin{equation}
c_{\gamma\square} = \frac{1}{g_2^2 - g_1^2} \lq 2g_2^2 c_{Z\square}  + 
(g_2^2 + g_1^2) c_{ZZ} - e^2 c_{\gamma\gamma} - (g_2^2 - g_1^2) c_{Z\gamma}\rq.
\label{eqn:hvvAN}
\end{equation}
For the sake of generality, we include CP-odd couplings parametrized by
${\tilde c}_{VV}$ in our calculation. Note that if one assumes that
the Higgs particle is a pure CP-even eigenstate, CP-odd operators are not
allowed\footnote{Present data are consistent with the CP-even nature of 
Higgs, however, it does allow a room for accommodating a small CP-violating 
component in the couplings of the Higgs boson with fermions and gauge
bosons~\cite{Dekens:2013zca, Agashe:2014kda, Dwivedi:2015nta,
  Cirigliano:2016njn}.}. 
The second class of operators is given by the anomalous contributions
to $Z\ell\ell$ vertex
\begin{equation}
\begin{split}
\lag^{Z\ell\ell}_{D=6} = \sqrt{g_1^2 + g_2^2} \, \sum_{\ell = e,\mu} Z_\mu \lq
\vphantom{I^3_{W,\ell}} \right.  
&\bar{\ell}_L \gamma^\mu  \( I^3_{W,\ell} - s_W^2 Q_\ell + \delta g_L^{Z\ell\ell} \) \ell_L + \\
+&\bar{\ell}_R \gamma^\mu  \( - s_W^2 Q_\ell + \delta g_R^{Z\ell\ell} \) \ell_R 
\left. \vphantom{I^3_{W,\ell}} \rq,
\label{eqn:zffAN}
\end{split}
\end{equation}
while the third class gives rise to $HV\ell\ell$ contact interactions
\begin{equation}
  \lag^{HZ\ell\ell}_{D=6} = 2\frac{\sqrt{g_1^2 + g_2^2}}{v} \sum_{\ell=e,\mu} \lq
  \delta g_L^{HZ\ell\ell}H Z_\mu \bar{\ell}_L \gamma^\mu \ell_L +
  \delta g_R^{HZ\ell\ell} H Z_\mu \bar{\ell}_R \gamma^\mu \ell_R\rq.
\label{eqn:hvllAN}
\end{equation}
If a linear realization of the
$SU(2)_L \otimes U(1)_Y$ symmetry is 
assumed, the contact terms in Eq.~(\ref{eqn:hvllAN}) are generated by
the same operators which give rise to vertex corrections in
Eq.~(\ref{eqn:zffAN}). In the Higgs basis they are set to be the same\footnote{In general, $\delta g_i^{Z\ell\ell}$ can have additional contributions
due to a redefinition of couplings and  
the choice of input parameter schemes.}, 
\begin{equation}
  \delta g_L^{Z\ell\ell} = \delta g_L^{HZ\ell\ell}, \qquad
  \delta g_R^{Z\ell\ell} = \delta g_R^{HZ\ell\ell}.
  \label{eqn:contrel}
\end{equation}
In this scenario, the coefficients for the contact
interactions are constrained by the EW precisions tests 
performed at LEP and their effects are expected to be rather small.
However, there are scenarios in which the coefficients in Eqs.~(\ref{eqn:zffAN}-\ref{eqn:hvllAN}) can be independent (see for instance Refs.~\cite{Isidori:2013cga, deFlorian:2016spz}). 
In this work we also assume flavor universality, so that
$g_{L,R}^{Zee}=g_{L,R}^{Z\mu\mu}$ 
and $g_{L,R}^{HZee}=g_{L,R}^{HZ\mu\mu}$. This assumption is very much consistent
with LEP data on $Z\ell\ell$ couplings. Any violation of this
assumption can be checked by 
comparing $H \to 2e2\mu$ with $H \to 4e$ and
$H \to 4\mu$~\cite{Gonzalez-Alonso:2015bha}. 

The last two contributions involve dipole
interactions between $Z$ bosons and leptons
and the dipole contact interactions
of the Higgs boson. These terms are proportional to lepton masses and
in the $m_l \to 0$ limit 
can be safely neglected. Moreover, as a consequence 
of the linearly realized electroweak symmetry in the 
$D = 6$ Lagrangian, the dipole parameters 
are proportional to the respective lepton dipole 
moments, which are tightly 
constrained by experimental data  
and usually neglected in the LHC analyses.
Note that a contact term involving $H$ 
and four leptons can only be generated by $D=8$ operators. One 
would be sensitive   
to such a contact term in the kinematic region where the $4\ell$
invariant mass is much  
higher than the Higgs mass which is not the case in the on-shell decay
of the Higgs boson. 

\section{Computational details}
\label{sec:calculation}
In order to study the possible BSM deviations in the Higgs decay into 
four charged leptons we have considered the effective
Lagrangian\footnote{This Lagrangian corresponds to  
  the {\it Beyond the Standard Model
    Characterization}~\cite{Falkowski:2015wza}
  one, restricted to
  the relevant operators for the $H\to 4\ell$ channels.}
\begin{equation}
  \lag_\mathrm{EFT} = \lag_\mathrm{SM} + \lag^{D=6}_{HVV} +
  \lag_{Z\ell\ell}^{D=6} + \lag_{HZ\ell\ell}^{D=6},
  \label{eqn:eftlagg}
\end{equation}
where $\lag_\mathrm{SM}$ is supplemented by the $D=6$
contributions in Eqs.~(\ref{eqn:Hvv})-(\ref{eqn:hvllAN}).
The master formula for the LO decay width,
in the presence of $D=6$ operators reads  
\begin{equation}
\Gamma_{\mathrm{LO}}^{D=6} \( H \to 4\ell\) = \frac{1}{2M_H} \int
\lg \left| \mat_\mathrm{SM} \right|^2 + 2\realpt(\mat_{D=6}\,\mat_\mathrm{SM}^*) +
\left| \mat_{D=6} \right|^2 \rg \rmd\Phi_4,
\label{eqn:h4l-dim6}  
\end{equation}

In addition to the anomalous part in the $HZZ$ and $Z\ell\bar{\ell}$ couplings,
the presence of $D=6$ operators gives rise to tree-level
$H\gamma\gamma$ and $HZ\gamma$ and $HZ\ell\bar{\ell}$ vertices
which are not present in the SM Lagrangian. 
The Feynman rules for these anomalous vertices
have been derived by implementing 
the effective Lagrangian
of  Eq.~(\ref{eqn:eftlagg}) in {\tt FeynRules
  2.0}~\cite{Alloul:2013bka}. For massless leptons we get,

\begin{equation}
  \begin{split}
    V_{H\gamma\gamma}^{\mu_1 \mu_2} (p_1,p_2) &= \frac{ie^2}{v} \lg \vphantom{\half} \right. c_{\gamma\gamma}
    \lq p_1^{\mu_2}p_2^{\mu_1} - \(p_1 \cdot p_2\) g^{\mu_1\mu_2}
    \rq  \\
    &+  \tilde{c}_{\gamma\gamma}~\epsilon_{\mu_1 \mu_2 p_1 p_2} \left.\vphantom{\half}\rg,
  \end{split}
  \label{eqn:hgg}
\end{equation}

\begin{equation}
  \begin{split}
    V_{HZ\gamma}^{\mu_1 \mu_2}  (p_1,p_2) = \frac{ie^2}{c_W s_W v}
    \lg \vphantom{\half} \right. &c_{Z\gamma}
    \lq p_1^{\mu_2}p_2^{\mu_1} - \(p_1 \cdot p_2\) 
    g^{\mu_1\mu_2}\rq  \\
      + &c_{\gamma\square} \lq p_2^2
    g^{\mu_1\mu_2}  \rq 
    + \tilde{c}_{Z\gamma}~\epsilon_{\mu_1 \mu_2 p_1 p_2}  \,
    \left.\vphantom{\half}\rg , 
    \label{eqn:hgz}
  \end{split}
\end{equation}

\begin{equation}
  \begin{split}
    V_{HZZ}^{\mu_1 \mu_2}  (p_1,p_2) = &+ \frac{i}{2} (g_1^2+g_2^2) v \( 1
    + \delta c_Z \) g^{\mu_1 \mu_2}
    + \frac{ie^2}{c_W^2 s_W^2 v} c_{ZZ}\lq p_1^{\mu_2}p_2^{\mu_1} -
    \(p_1 \cdot p_2\)g^{\mu_1\mu_2} \rq \\ 
     &+\frac{ie^2}{s_W^2 v} c_{Z\square}  \lq 
       \(p_1^2 +
     p_2^2\) g^{\mu_1 \mu_2} \rq 
    + \frac{ie^2}{c_W^2 s_W^2 v} 
    \tilde{c}_{ZZ}~\epsilon_{\mu_1 \mu_2 p_1 p_2},
    \label{eqn:hzz}
  \end{split}
\end{equation}

\begin{equation}
  V_{HZ\ell\bar{\ell}}^\mu = \frac{2ie}{c_W s_W v} \gamma^\mu
  \( \delta g_L^{HZ\ell\ell} \omega_{L} + \delta g_R^{HZ\ell\ell}
  \omega_{R} \),
  \label{eqn:hzll}
\end{equation}

\begin{equation}
  V_{Z\ell\bar{\ell}}^\mu = i\sqrt{g_1^2+g_2^2} \gamma^\mu \lq \( g^{Z\ell\ell}_L +
  \delta g_L^{Z\ell\ell} \) \omega_{L} + 
                                              \( g^{Z\ell\ell}_R +
  \delta g_R^{Z\ell\ell}  \) \omega_{R} \rq
    \label{eqn:hll}
\end{equation}
where, $\epsilon_{\mu \nu p_i p_j} =
\epsilon_{\mu\nu\rho\sigma}p^\rho_i p^\sigma_j$, $g^{Z\ell\ell}_L = I^3_{W,\ell} - g_1^2/(g_1^2+g_2^2) Q_\ell $ and
$g^{Z\ell\ell}_R = - g_1^2/(g_1^2+g_2^2) Q_\ell$. 
In the previous expressions $p_1$ and $p_2$ are the incoming momenta
of gauge bosons and $\omega_{L,R} = \frac{1}{2} ( 1 \mp \gamma_5 )$.
The calculation of new matrix elements for $H \to 2e2\mu$ and $H
\to 4e/4\mu$  
has been carried out by means of the symbolic 
manipulation program {\tt FORM}~\cite{Vermaseren:2000nd}, and they
have been included in a new version of the code {\tt Hto4l}, which is
publicly available. As in other Monte Carlo tools for Higgs physics, such as
{\tt HiGlu}~\cite{Spira:1995mt, Spira:1996if}, {\tt
  Hawk}~\cite{Ciccolini:2007jr, Ciccolini:2007ec, Denner:2011id,
  Denner:2014cla} and {\tt HPair}~\cite{Dawson:1998py, Grober:2015cwa}, the 
new version of {\tt Hto4l} provides the possibility to compare
present and future Higgs data with theoretical predictions derived in
an EFT context.

Since we have neglected the lepton masses, the 
matrix elements for $4e$ and $4\mu$ are the same. 
As a consistency check we have compared 
the value of the matrix elements implemented in {\tt Hto4l} with the
ones generated with {\tt MadGraph5@MC\_NLO}~\cite{Alwall:2014hca} for
several phase-space points, finding excellent agreement.

Few important remarks are in order: first of all we note that
the quadratic part $\left| \mat_{D=6} \right|^2$ of Eq.~(\ref{eqn:h4l-dim6})
is suppressed by a factor $1/\Lambda^4$. From the point of view
of the EFT expansion, it contributes at the same level of $D=8$
operators. Moreover, different bases of $D=6$ operators are 
equivalent only at the order of $1/\Lambda^2$ and they differ by terms 
which are of order $1/\Lambda^4$. It follows that predictions 
obtained by using only $D=6$ operators are not complete at 
the ${\cal O}(1/\Lambda^4)$. 
There are different approaches in the literature regarding the treatment 
of quadratic contributions in the analyses. One approach consists in making
linear approximation for the theoretical predictions and including the quadratic
contributions in the estimation of the theoretical uncertainty.
In this context, the constraints derived in one basis can be translated to other bases.
Another approach keeps always the quadratic contributions in the
calculations. The latter improves the accuracy of the calculation
 whenever the contribution of D=8 operators is subdominant~\cite{Contino:2016jqw}.
Pragmatically, we have included the quadratic contributions in our
calculation with the possibility of switching them on and off in the code.

In order to guarantee flexibility in the choice
of the basis, 
a provision of calculating $H \to 4\ell$ matrix elements in SILH and Warsaw 
bases which are not affected by the basic assumptions of the Higgs basis, 
is also made. For that a separate dictionary between the {\it
  anomalous coupling parameters}  
appearing in the Feynman rules (Eqs.~\ref{eqn:hgg}-\ref{eqn:hll}) and the 
Wilson coefficients of the SILH and Warsaw bases is implemented in the
code and it is listed in  
Appendix~\ref{appen:dictionary}.

\section{Numerical results}


In this section we present some numerical results obtained with the
new version of {\tt Hto4l} for the $H \to 4\ell$ decay channel in
the presence of $D=6$ operators of the Higgs basis. The
results have been obtained with the same SM
input parameters as in Ref.~\cite{Boselli:2015aha}. 
In the Higgs basis, the \{$G_F,\alpha,M_Z$\} input parameter
scheme is assumed. A shift to the \{$G_F,M_W,M_Z$\} input parameter 
scheme, which we have adopted, introduces corrections proportional to $\delta m$ (see Eq.~\ref{eqn:deltam}) 
in couplings and parameters dependent on the input parameters.      

Since the anomalous vertices $V_{H\gamma\gamma}$ and $V_{HZ\gamma}$ 
enter the calculation of the $H \to 4\ell$ partial decay width
we expect an important BSM
contribution coming from the
kinematic configurations with one of the lepton pair invariant masses 
close to zero. In order to get rid of these
contributions which would be hardly accessible by the
experiments, we have implemented a lower cut of 15 GeV on the leading
and subleading same-flavor opposite-sign (SFOS) lepton pair invariant masses. 

\subsection{BSM predictions for the partial decay width}


The modification of the $H \to 2e2\mu$ decay width in the presence of the 
CP-even and CP-odd parameters of the Higgs basis can be parametrized 
as,
\begin{eqnarray}
  R_{\mathrm{BSM}} &=& \frac{\Gamma_{\mathrm{BSM}}}{\Gamma_{\mathrm{SM}}} (H \to 2e2\mu) \nonumber \\ 
             &=& 1.00 + \sum_i X_i c_i + \sum_{ij} X_{ij}c_i c_j 
  + \sum_{ij} {\tilde X}_{ij}{\tilde c}_i {\tilde c}_j.
  \label{eqn:GGG}
\end{eqnarray}
where $c_i = \{\delta c_Z, c_{\gamma\gamma}, c_{Z\gamma}, c_{ZZ},
c_{Z\square}, \delta g^{Z\ell\ell}_L,  
\delta g^{Z\ell\ell}_R, \delta g^{HZ\ell\ell}_L, \delta g^{HZ\ell\ell}_R \}$ and 
${\tilde c}_i = \{ {\tilde c}_{\gamma\gamma}, {\tilde c}_{Z\gamma}, {\tilde c}_{ZZ} \}$. 
The absence of linear terms in CP-odd parameters is related to the
fact that the partial decay width is a CP-even quantity.
The coefficients of the linear and quadratic terms are given
by\footnote{The values of the coefficients depend on the choice of
  the SM input parameters and on the selection cuts.},

\begin{equation*}
  X_i = 
  \begin{pmatrix}
    2.00 & 0.0115 & 0.170 & -0.232 & 0.301 & -8.77 & 7.04 & 4.47 & -3.58
  \end{pmatrix},
\end{equation*}

\begin{equation*}
  X_{ij} =
  \begin{pmatrix}
    1.00 & 0.0115 & 0.170 & -0.232 & 0.301 & -8.77 & 7.04 & 4.47 & -3.58 \\
    0 & 0.055 & 0.0706 & -0.0312 & -0.0448 & -0.227 & -0.179 & -0.181 & 0.174 \\
    0 & 0 & 0.768 & -0.490 & -0.702 & -3.47 & -2.80 & 2.81 & 2.740 \\
    0 & 0 & 0 & 0.114 & 0.273 & 2.23 & 0.696 & -1.55 & -0.873 \\
    0 & 0 & 0 & 0 & 0.265 & 0.566 & 3.41 & -0.974 & -2.51 \\
    0 & 0 & 0 & 0 & 0 & 25.4 & -15.4 & -25.9 & 7.85 \\
    0 & 0 & 0 & 0 & 0 & 0 & 22.0 & 7.85 & -22.4 \\
    0 & 0 & 0 & 0 & 0 & 0 & 0 & 7.85 & -1.58 \\
    0 & 0 & 0 & 0 & 0 & 0 & 0 & 0 & 7.50
  \end{pmatrix},
\end{equation*}

\begin{equation*}
  {\tilde X}_{ij} = 
  \begin{pmatrix}
    0.0487 & -0.00745 & 0.0000910 \\
    0 & 0.308 & -0.00592 \\
    0 & 0 & 0.00317
  \end{pmatrix}.
\end{equation*}

\begin{figure}
  \includegraphics[width = 0.5 \textwidth]{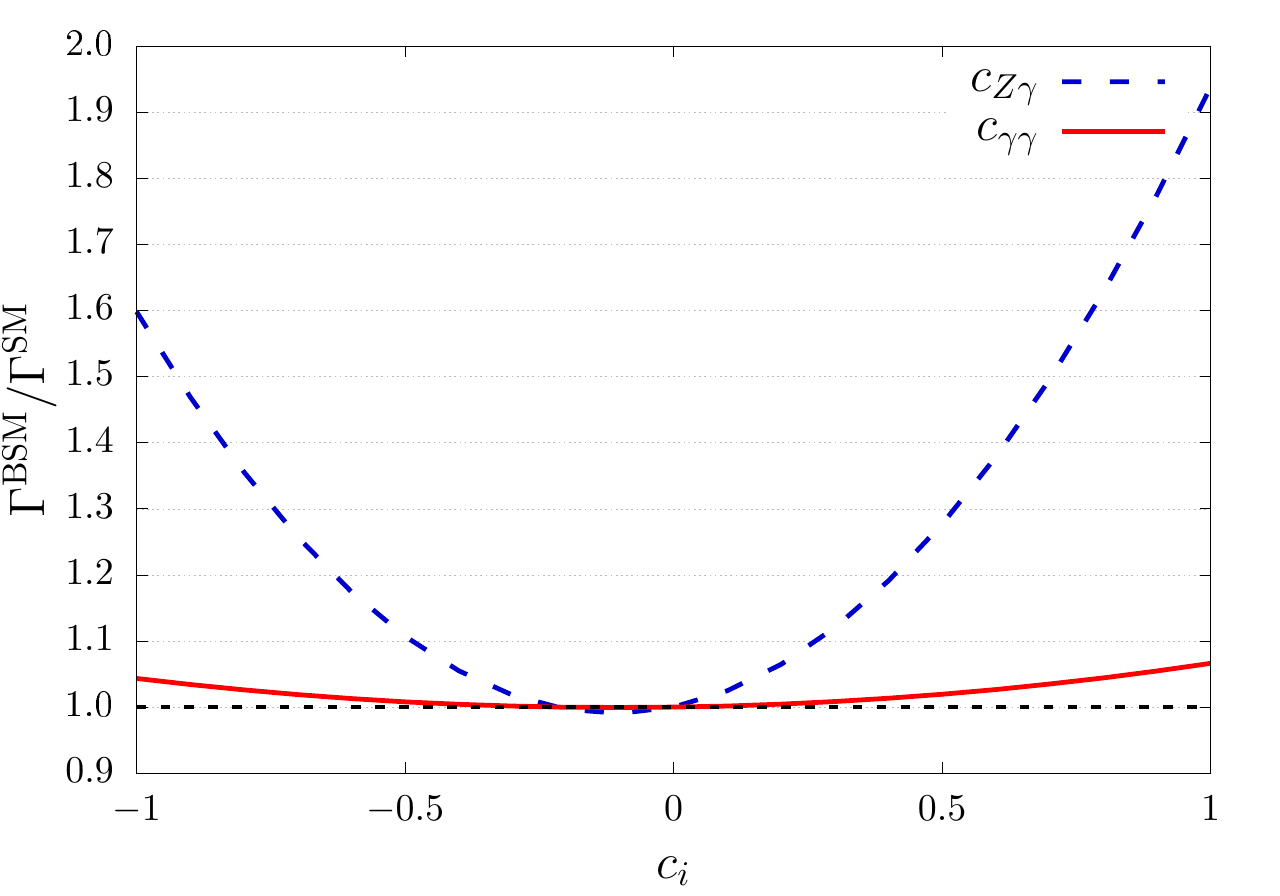}
  \includegraphics[width = 0.5 \textwidth]{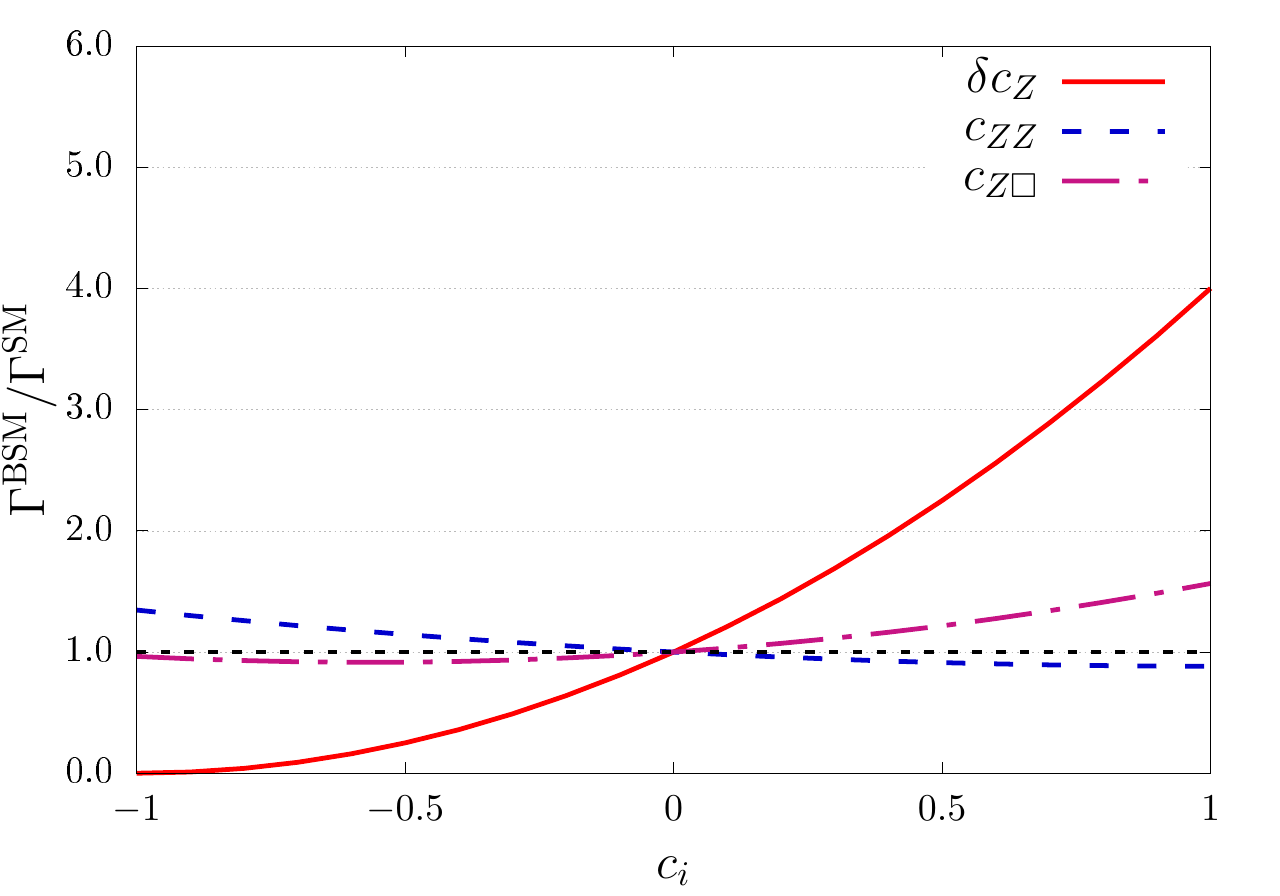}
  \includegraphics[width = 0.5 \textwidth]{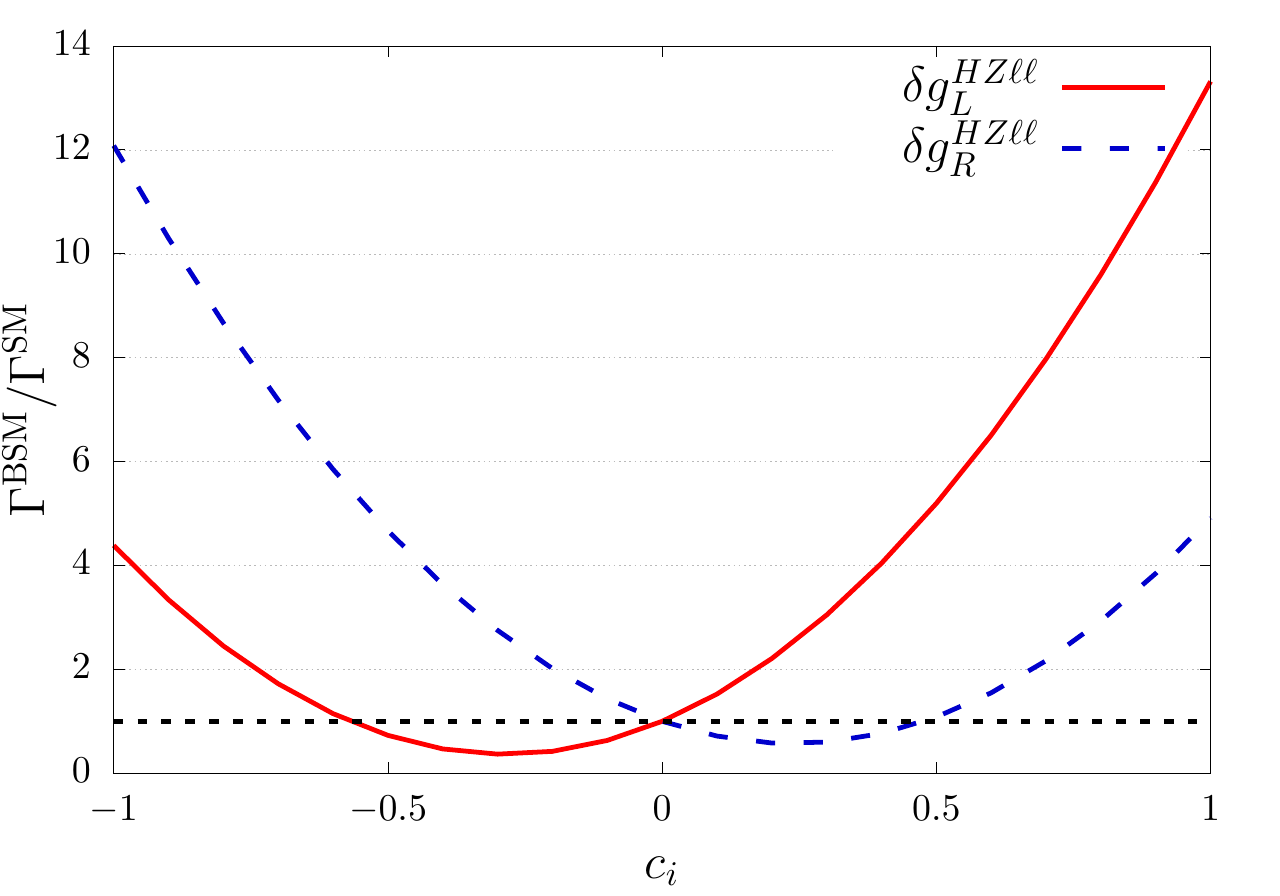}
  \includegraphics[width = 0.5 \textwidth]{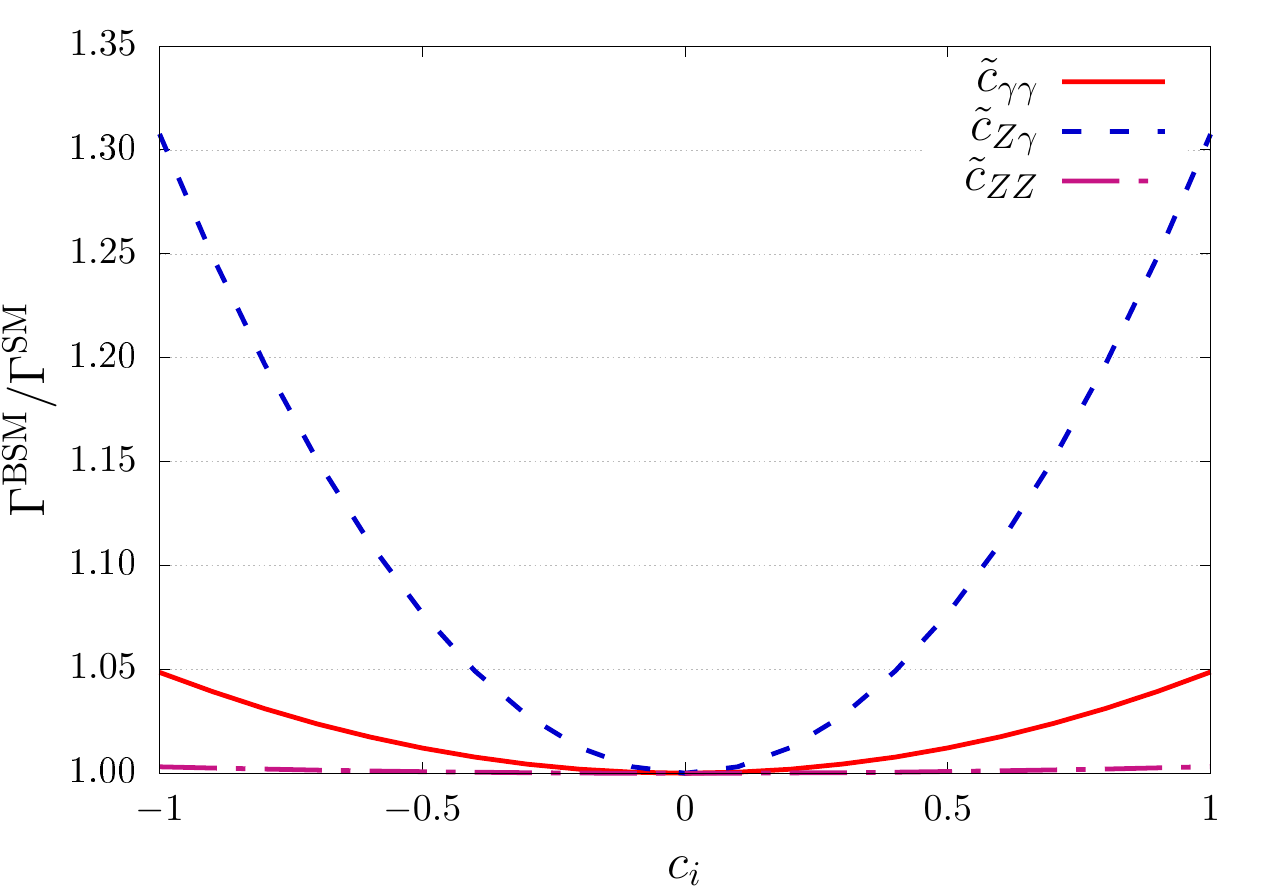} 
 \caption{\footnotesize Modifications of the $H \to 2e2\mu$ decay width in
   the presence of Higgs basis parameters. In the left plot on the lower line, the
   coefficients of the contact interactions are taken independent of
   the anomalous $Z\ell\ell$ interactions.} 
 \label{fig:scan-param}
\end{figure}

The corresponding coefficients for $H \to 4e$ are given in
Appendix~\ref{appen:H4e}.  
Note that in the above we have intentionally kept $\delta g^{HZ\ell\ell}_i $ 
and $\delta g^{Z\ell\ell}_i $ independent of each other to cover 
the scenario in which new physics parametrization leads to additional
contributions in $\delta g^{Z\ell\ell}_i$.
{ In the Higgs basis we must set 
$\delta g^{HZ\ell\ell}_i = \delta g^{Z\ell\ell}_i$.} The relative 
importance of various parameters of the Higgs basis in modifying 
the partial decay width can be inferred 
from the size of the coefficients derived above.  
To illustrate the relative effect of the parameters more clearly, 
in Fig.~\ref{fig:scan-param}   
we plot the ratio in Eq.~(\ref{eqn:GGG}) 
by scanning each parameter in the range 
between -1.0 and +1.0. 
Among CP-even parameters related to the
$HVV (V=\gamma,Z)$ couplings, the change in partial decay width due to
$c_{\gamma\gamma}$ is the smallest, while $\delta c_Z$, which gives
rise to a
SM-like anomalous coupling, changes the width maximally. Due to different
propagator  
effects the effect of $c_{Z\gamma}$ is larger than that of $c_{\gamma\gamma}$.
The contact interaction parameters, however, modify the 
width the most because of no propagator suppression.

In the CP-even case, these scan plots display the importance of the linear
terms with respect to the 
quadratic terms. For instance, we find that for 
$c_{\gamma \gamma}$ and $c_{Z\gamma}$ the quadratic
contributions dominate over the linear ones in most of the parameter space, leading 
to an overall enhancement of the decay width. On the other hand, 
for $\delta c_Z, c_{ZZ}$ and $c_{Z\square}$, the linear 
terms play an important role and the decay width can become smaller than 
its SM value in certain regions of parameter space. Also, the effect of  $c_{ZZ}$ and 
$c_{Z\square}$ on the partial width is opposite in nature. For contact interaction  
parameters the quadratic terms dominate over the linear ones, except
for a small region of parameter space between 0 and 0.5 (-0.5) for $\delta g^{HZ\ell\ell}_L$
($\delta g^{HZ\ell\ell}_R$) where the ratio goes below 1. 

As mentioned before, the CP-odd
parameters contribute to the total Higgs decay rate only at the
quadratic level leading to the ratio always greater than 1. Among the CP-odd 
parameters, the change of the decay width due to ${\tilde c}_{Z\gamma}$ is
the largest one 
while the corresponding change due to ${\tilde c}_{ZZ}$ is the
smallest one.
Information on CP-odd linear terms can be 
accessed from specific kinematic distributions which we discuss later.

It is important to stress that some of these parameters are already
constrained by the available experimental data from LEP and
LHC. For instance, by using LHC Run-I data~\cite{Falkowski:2015fla},
$c_{\gamma\gamma}$ and $c_{Z\gamma}$ are constrained respectively at
the $10^{-3}$ and $10^{-2}$ level.  
On the contrary $\delta c_Z$, $c_{ZZ}$ and $c_{Z\square}$ are
loosely constrained. 
An approximate degeneracy, which corresponds to a strong correlation, is found between 
$c_{ZZ}$ and $c_{Z\square}$ ($\rho_{ij}=-0.997$). Including the LEP data on $WW$ production, 
$\delta c_Z$ and  $c_{Z\gamma}$ become more constrained and the flat
direction between  
$c_{ZZ}$ and $c_{Z\square}$ is also lifted to some
extent ($\rho_{ij}=-0.96$) ~\cite{Falkowski:2015jaa}.  
These conclusions assume linear dependence of Higgs signal strength
observables on parameters. 

It has been argued that there is no model independent constraint on $c_{ZZ}$
and $c_{Z\square}$ because including contributions which are
quadratic in these parameters would dramatically change the corresponding
best-fit values and the relative uncertainties. To this end, more data
are needed and the complementary information coming from kinematic
observables will be helpful to improve the constraints on these 
coefficients~\cite{DiVita:2017eyz, Durieux:2017rsg}. 

Furthermore, the couplings of the $Z$ boson
to charged leptons are constrained by considering Electroweak
precision data (see Refs.~\cite{Berthier:2015gja, Berthier:2015oma}
for recent analyses where SMEFT theoretical errors are taken into
account).
In our framework these constraints  
are also applicable to the parameters of the $ZH\ell\ell$ contact
interactions.
To obtain any constraint on the CP-odd parameters, it is necessary to
go beyond the linear approximation for Higgs observables.
Interpreting the results obtained on CP-odd
parameters of the SILH basis in Ref.~\cite{Ferreira:2016jea} using
current Higgs data, we find that ${\tilde 
  c}_{\gamma\gamma}$  
is constrained at 1\% level. However, the allowed values for |${\tilde
  c}_{Z\gamma}$| and  
|${\tilde c}_{ZZ}$| can be as large as 0.7 and 0.5 respectively. In
the following we focus  
on the parameters which are loosely constrained by the data and have
non-negligible effects  
on partial decay width.

\subsection{BSM predictions for kinematic distributions}
\begin{figure}[ht]
   \includegraphics[width = .88 \textwidth]{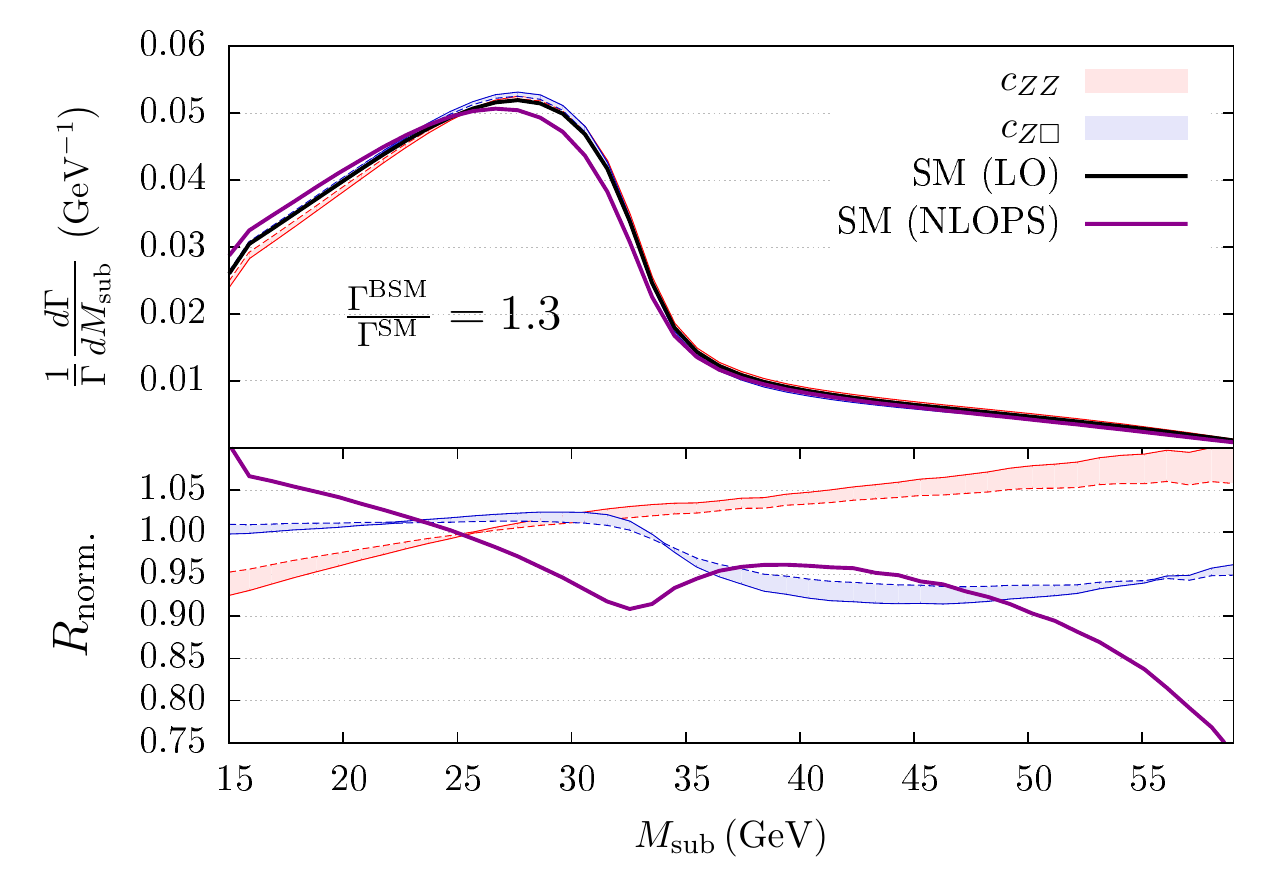}   
   \includegraphics[width = .88 \textwidth]{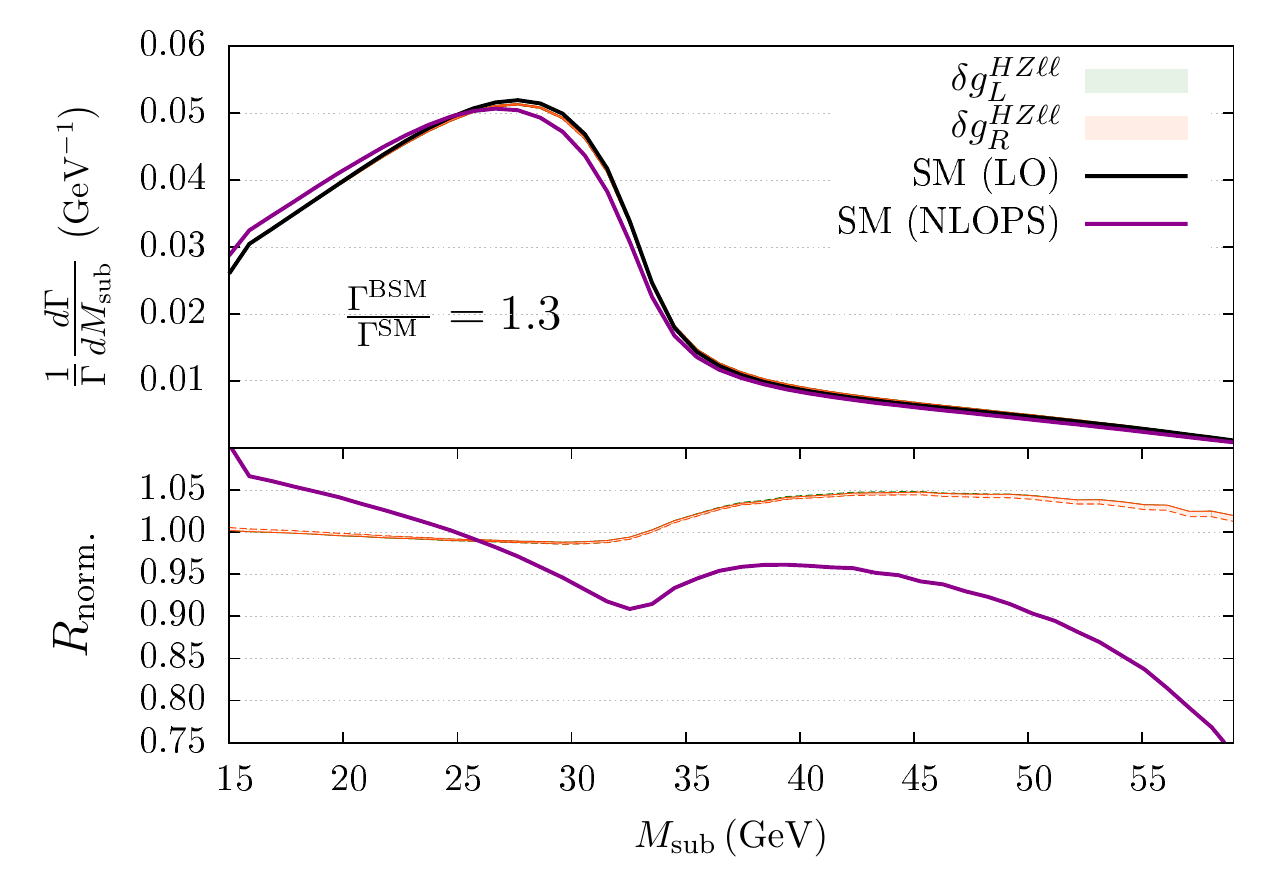}  
\caption{\footnotesize{Upper panels: normalized distributions for
    $M_\mathrm{sub}$ at LO (black curve), at NLOPS EW accuracy
    (violet curve) and for values of $c_{ZZ}$ (red curves), $c_{Z\square}$
    (blue curves), $\delta g^{HZ\ell\ell}_l$ and $\delta
    g^{HZ\ell\ell}_R$  giving rise to a 
    ratio $R_{\mathrm{BSM}}$ = 1.3.   
    Lower panels: Normalized
    ratios, according to the definitions in Eq.~(\ref{eqn:rn}). Solid
    lines refer to pure 
    $1/\Lambda^2$ effects, for dashed lines quadratic
    contributions are also included. The shaded bands between solid and dashed lines 
          highlight the shape variations induced by the inclusion of quadratic terms.}} 
 \label{fig:msub}
 \end{figure}

 \begin{figure}[ht]
   \includegraphics[width = .88 \textwidth]{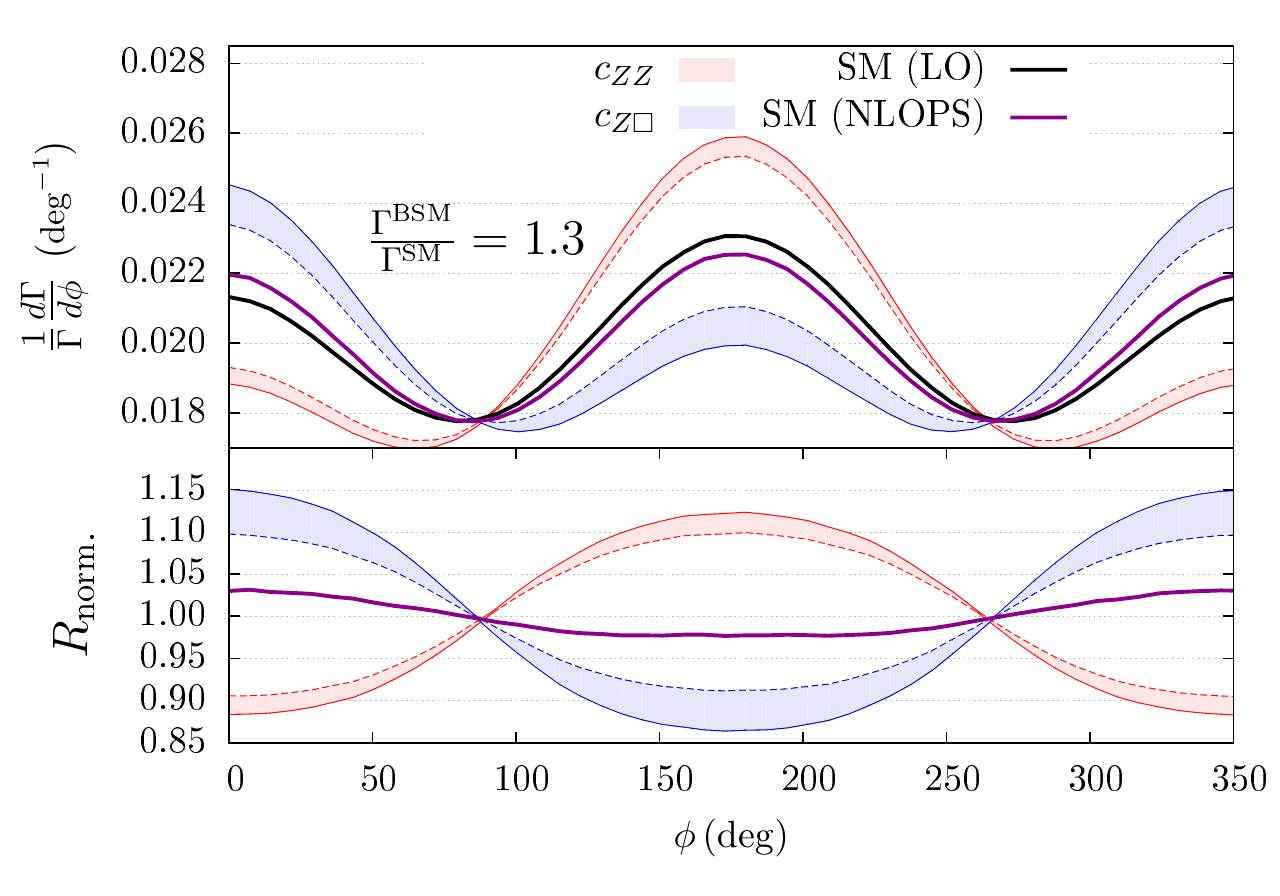}   
   \includegraphics[width = .88 \textwidth]{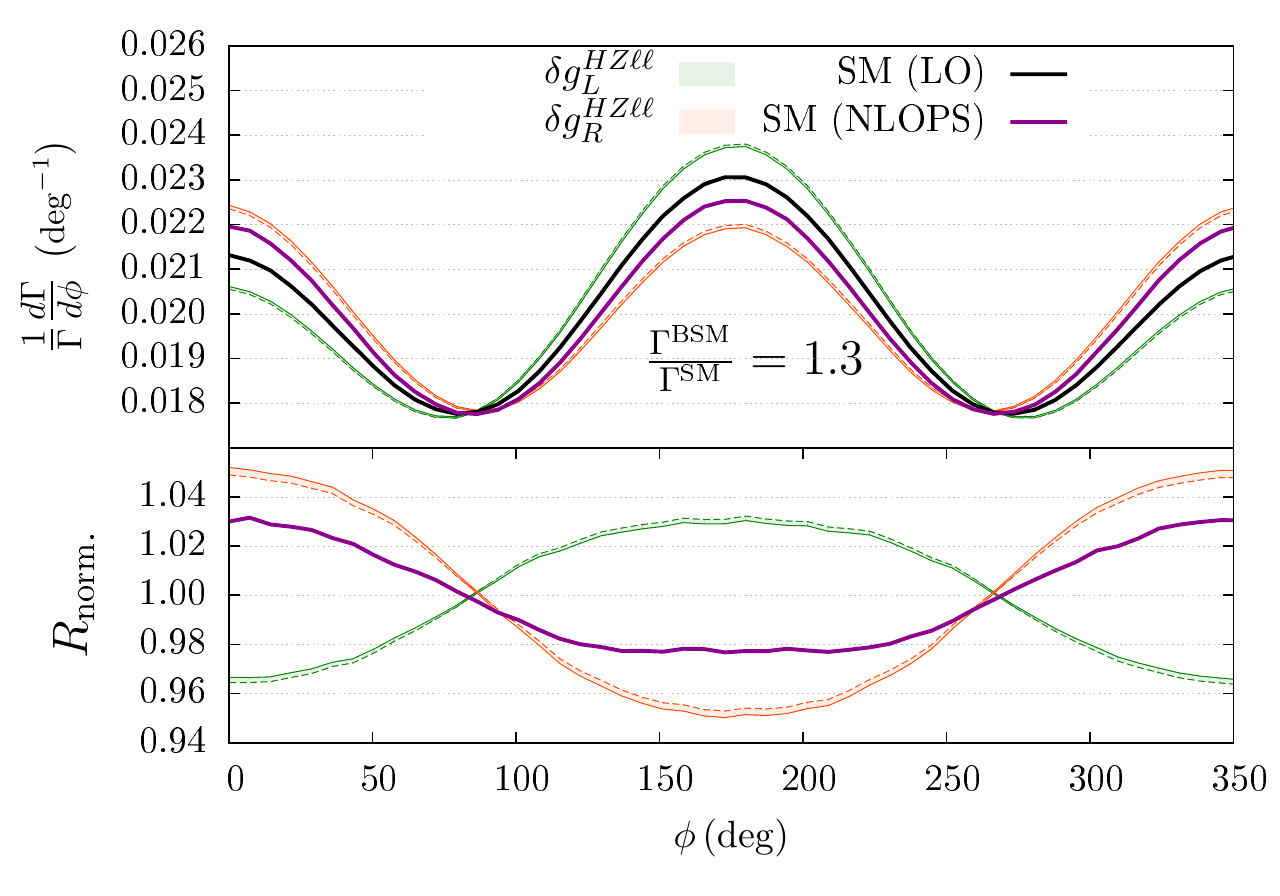}  
   \caption{\footnotesize{Upper panels: normalized distributions for
       $\phi$ at LO (black curve), at NLOPS EW accuracy
       (violet curve) and for values of $c_{ZZ}$ (red curves), $c_{Z\square}$
       (blue curves), $\delta g^{HZ\ell\ell}_l$ and $\delta
       g^{HZ\ell\ell}_R$  giving rise to a 
       ratio $R_{\mathrm{BSM}}$ = 1.3.  
       Lower panels: Normalized ratios, according to the definitions in Eq.~(\ref{eqn:rn}). Solid
       lines refer to pure 
       $1/\Lambda^2$ effects, for dashed lines quadratic
       contributions are also included. The shaded bands between solid and dashed lines highlight the shape variations induced by the inclusion of quadratic terms.}}
 \label{fig:phi1}
 \end{figure}

  \begin{figure}[ht]
   \includegraphics[width = .88 \textwidth]{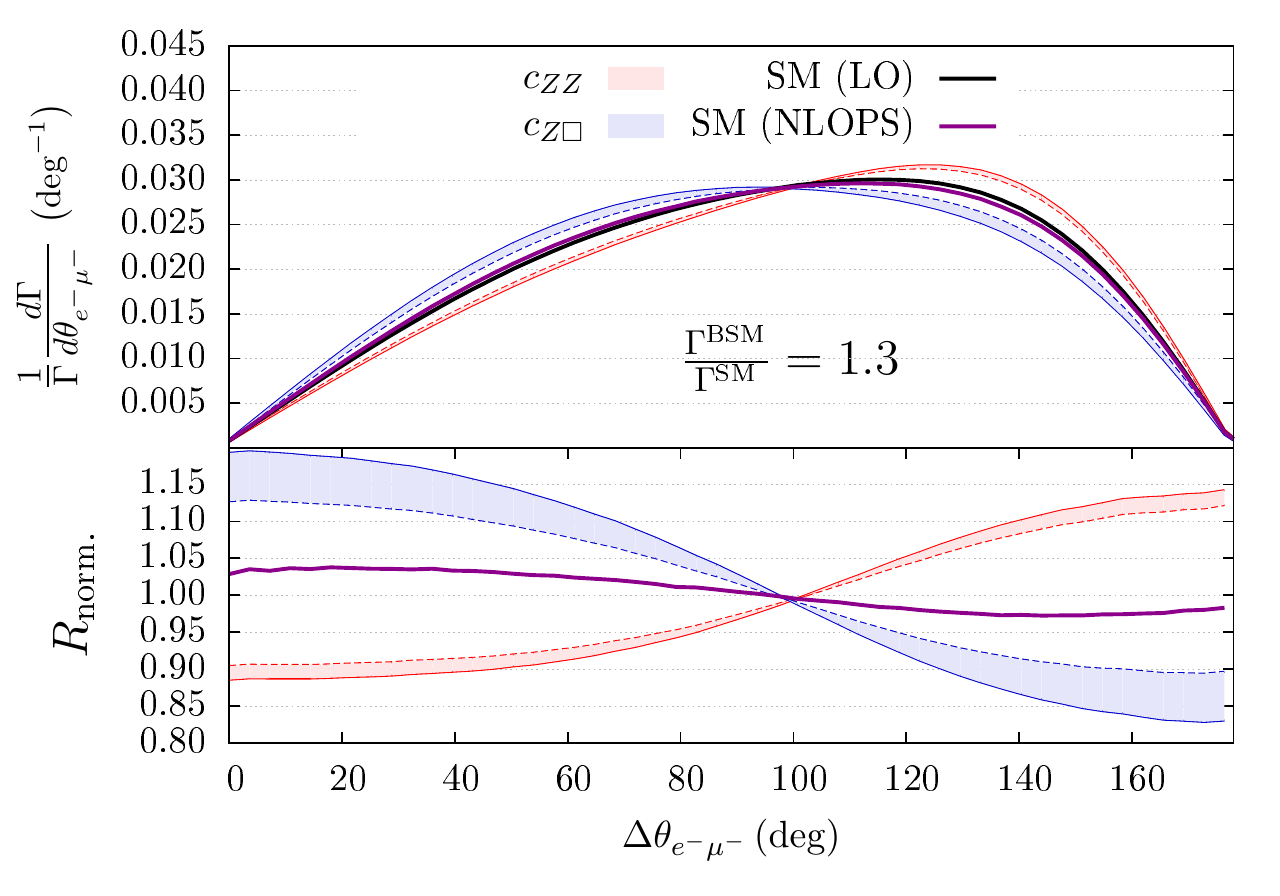}   
   \includegraphics[width = .88 \textwidth]{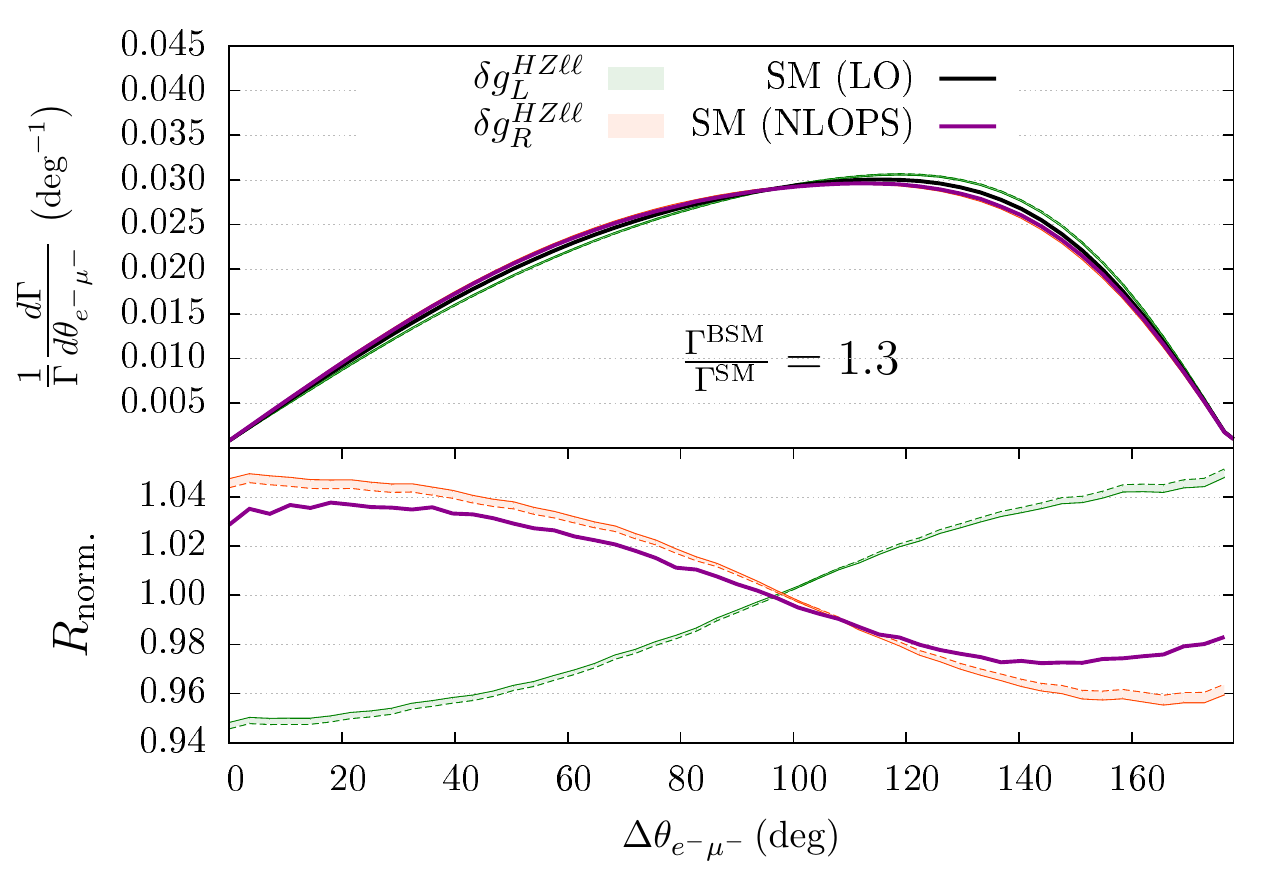}  
\caption{\footnotesize{Upper panels: normalized distributions for
    $\Delta\theta_{e^-\mu^-}$ at LO (black curve), at NLOPS EW accuracy
    (violet curve) and for values of $c_{ZZ}$ (red curves), $c_{Z\square}$
    (blue curves), $\delta g^{HZ\ell\ell}_l$ and $\delta
    g^{HZ\ell\ell}_R$  giving rise to a 
    ratio $R_{\mathrm{BSM}}$ = 1.3. 
Lower panels: Normalized
    ratios, according to the definitions in Eq.~(\ref{eqn:rn}). Solid
    lines refer to pure 
    $1/\Lambda^2$ effects, for dashed lines quadratic
    contributions are also included. The shaded bands between solid and dashed lines 
          highlight the shape variations induced by the inclusion of quadratic terms.}}
 \label{fig:cos}
  \end{figure}

In this section we use the new version of {\tt Hto4l}
to simulate the decay of the Higgs boson into four charged
leptons in the presence of $D=6$ operators at the differential level. 
The study of distributions can provide
complementary information to the analyses of signal
strengths and BRs. For the sake of simplicity we consider one
parameter at a time while  
the remaining ones are artificially set to zero. More sophisticated
analyses, where correlations among various coefficients are taken
into account, are beyond the 
scope of this article.

\begin{table}
    \begin{center}
    \begin{tabular}{c | c | c }
      $c_i$ & int. & quad. \\
      \hline
      $c_{ZZ}$ & -1.29 & -0.897 \\
      $c_{Z\square}$ & 0.996 & 0.638 \\
      $\delta g^{HZ\ell\ell}_L$ &  0.067 & 0.060 \\
      $\delta g^{HZ\ell\ell}_R$ & -0.084 & -0.073 \\
      ${\tilde c}_{Z\gamma}$ & 0 & $\pm 1.0$
    \end{tabular}
    \caption{\footnotesize{Values of the Higgs basis parameters 
    which modify the $H \to 2e2\mu$ partial decay width by about 30\%. 
    These values are derived considering linear (int.) and quadratic (quad.) 
    dependence on the parameters.
    For the quadratic case, out of two possible values we choose the
    smaller ones. }} 
    \label{tab:values}
    \end{center}
\end{table}

The parameters of our interest are $c_{ZZ}$, $c_{Z\square}$ and
${\tilde c}_{Z\gamma}$. Moreover, since $H \to 4\ell$ decay can provide
information on the contact $ZH\ell\ell$  
interaction, we will also consider the effect of $\delta
g^{HZ\ell\ell}_i$ independent  
of $\delta g^{Z\ell\ell}_i$. To emphasize the characteristic effects
of these parameters on distributions, we consider a scenario in which
the parameters lead to the same deviation in partial decay width. In particular, 
we choose the benchmark values for these parameters by  
considering an excess of 30\% in $\Gamma^{\mathrm{BSM}}( H \to
2e2\mu)$. In table~\ref{tab:values},  
the benchmark values are reported by keeping only the interference
terms and also by including the quadratic terms in the calculation.

Among the observables taken into account, the most sensitive ones to
BSM kinematic effects turn out to be  
\begin{itemize}
\item the subleading 
  lepton pair invariant mass $M_\mathrm{sub}$\footnote{
    The leading 
    lepton pair invariant mass is defined as the 
    SFOS lepton pair invariant mass closest
    to the $Z$ boson mass.}; 
\item the angle $\phi$  between the decay planes of the two
  intermediate gauge-bosons in the Higgs rest-frame;
\item the angle $\Delta\theta_{e^-\mu^-}$  between the electron and
  the muon in the Higgs rest-frame.
\end{itemize}
 
\begin{figure}[ht]
  \includegraphics[width = .88 \textwidth]{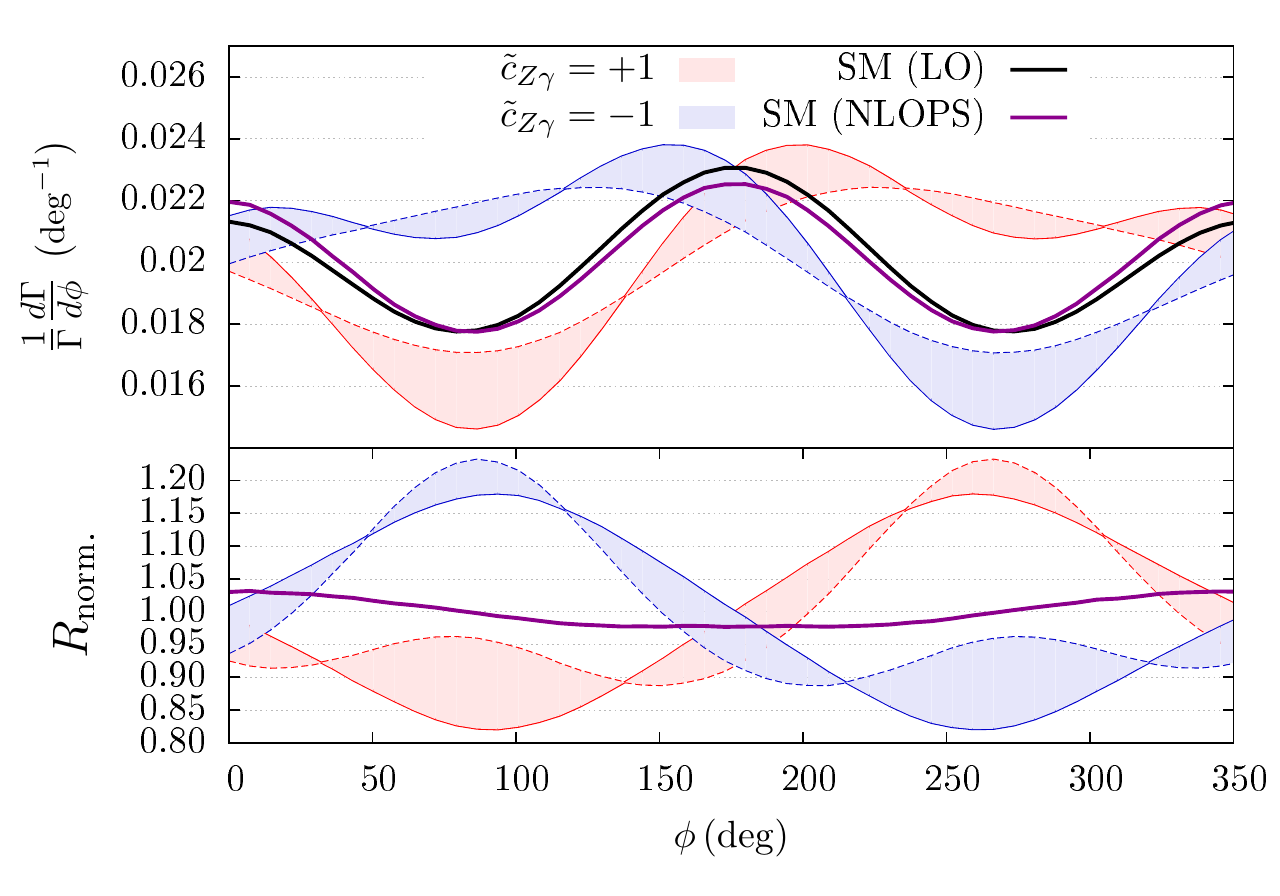}   
   \caption{\footnotesize{Upper panels: normalized distributions for
       $\phi$ at LO (black curve), at NLOPS EW accuracy
       (violet curve) and for values of ${\tilde c}_{Z\gamma}=+1$ (red
       curves), ${\tilde c}_{Z\gamma}=-1$.  
       Lower panels: Normalized ratios, according to the definitions in
       Eq.~(\ref{eqn:rn}). Solid
       lines refer to pure 
       $1/\Lambda^2$ effects, for dashed lines quadratic
       contributions are also included. The shaded bands between solid and dashed lines 
          highlight the shape variations induced by the inclusion of quadratic terms.}}
\label{fig:tCza}
\end{figure}

In Figs.~\ref{fig:msub}-\ref{fig:tCza}, we compare the BSM predictions for the
normalized distributions of these observables with the SM ones at
Leading Order (LO) and at Next-to-Leading Order EW accuracy matched to
a QED Parton Shower (NLOPS in the following), 
{\it i.e.} the highest SM theoretical accuracy achievable with
{\tt Hto4l}.
In order to
better highlight the kinematic effects we also plot the 
normalized ratios $R_{\textrm{norm.}}$, defined as:
\begin{equation}
  R_{\textrm{norm.}} = \frac{\frac{1}{\Gamma^{(i)}}\frac{d
      \Gamma^{(i)}}{d X}}{\frac{1}{\Gamma^{\mathrm{LO}}}\frac{d
      \Gamma^{\mathrm{LO}}}{d X}},
  \label{eqn:rn}
\end{equation}
where $X$ is a generic observable, while $i = c_i$ or $i = \mathrm{NLOPS}$. 
Note that to calculate the BSM excess in each bin this ratio has to 
be multiplied by 1.3.
Continuous lines in the plots refer to distributions obtained by
considering only the effects of interference, while for the dashed ones
quadratics effects have been also taken into account. 
Several remarks are in order:

\begin{itemize}
\item The angular
  variables turn out to be more sensitive to BSM kinematic effects
  than $M_\mathrm{sub}$. 
  
\item Among the CP-even parameters considered in the analysis,
  $c_{ZZ}$ and $c_{Z\square}$ have a larger impact on the normalized
  distributions than $\delta g^{HZ\ell\ell}_{L,R}$ (Figs.~\ref{fig:msub}-\ref{fig:cos}).
  As far as $\phi$ and $\Delta\theta_{e^-\mu^-}$ are concerned, the BSM
  effects are larger than the SM higher order corrections, while the
  effects of contact interactions seem to be of the same order of
  magnitude as EW corrections~(Figs.~\ref{fig:phi1}-\ref{fig:cos}).
  
\item The effect of
  $c_{ZZ}$ on $M_{\mathrm{sub}}$ monotonically increases as we go
  towards the tail of the distribution reaching an excess close to
  40\%. In case of $c_{Z\square}$, the ratio grows mildly in the
  beginning and starts decreasing beyond 33 GeV~(upper panel plot in Fig.~\ref{fig:msub}). 
  
 \item The effects of $c_{ZZ}$ and $c_{Z\square}$ on angular observables
 are opposite in nature. 
  In the presence of
  $c_{ZZ}$ more events fall in the central $\phi$ region, while in
  the presence of $c_{Z\square}$, the edges get more populated~(upper panel plot in Fig.~\ref{fig:phi1}). Similarly,
  looking at the $\Delta\theta_{e^-\mu^-}$ distributions, we find that
  $c_{ZZ}$, unlike $c_{Z\square}$, puts more events in the region where the angle between
  electron and muon is greater than 90 degrees~(upper panel plot in Fig.~\ref{fig:cos}). 
  
\item For $c_{ZZ}$ and $c_{Z\square}$ the effects of quadratic terms
  depends on the considered observable and in general are not
  negligible. The difference between predictions obtained
  by including only $1/\Lambda^2$ terms, with respect to those
  including also the quadratic contributions, turns out to be larger for
  the angular observables than on $M_{\mathrm{sub}}$~(upper panel plots in 
  Figs.~\ref{fig:msub}-\ref{fig:cos}). For instance, as
  far as $c_{Z\square}$ is concerned, the quadratic contributions can give
  up to a further 5\% difference at the level of normalized ratios in
  some of the bins.
  
  \item On $M_{\mathrm{sub}}$, the effects of $\delta g^{HZ\ell\ell}_{L}$ and 
  $\delta g^{HZ\ell\ell}_{R}$ are the same (lower panel plot in
    Fig.~\ref{fig:msub}). However, the angular observables  
  can be used to discriminate the two parameters (lower panel plots in
  Figs.~\ref{fig:phi1}-\ref{fig:cos}). Since the 
  interference and quadratic values   
  obtained for them are small and close to each other, the
  contribution of quadratic 
  terms over the linear one is very minute.
  
\item We find that for our choice of values for ${\tilde c_{Z\gamma}}$, 
 $\phi$ is the most sensitive observable. The angle 
   $\phi$ is a CP-odd observable and, unlike the partial decay width, it is
  sensitive to the linear term in ${\tilde c_{Z\gamma}}$. Also, 
  for the same reason, it can provide information on the sign 
  of the parameter. These features are 
  clearly visible in Fig.~\ref{fig:tCza}.

\end{itemize}

\subsection{Future prospects at HL-LHC}

One of the main opportunities of the HL-LHC
program is to enable precise measurements of the Higgs boson properties, 
such as the presence of anomalous couplings to bosons and fermions.
It has been shown that kinematic
distributions, such as the $p_T$ of the Higgs boson, can significantly
improve the multi-dimensional parameter fit~\cite{Englert:2015hrx}.
In this section we present the results of a $\chi^2$ analysis carried
out in the context of the High-Luminosity LHC (HL-LHC). The study has
the illustrative purpose to assess how $H\to 2\ell 2\ell'$ angular
observables can be exploited to constrain SMEFT coefficients in future
analyses of LHC data (see Ref~\cite{DiVita:2017eyz, Durieux:2017rsg}).  
Due to the large limits and to the
strong correlation arising from current constraints, the 
analysis is restricted to the $c_{ZZ}-c_{Z\square}$.
At LHC, the $H\to 4\ell$ decay has been observed mainly in the gluon-gluon fusion channel.
The observed signal strength using the 7 and 8 TeV LHC data 
is given by $\mu^{4\ell}_{gg{\rm F}} = 1.13^{+0.34}_{-0.31}$~\cite{Khachatryan:2016vau}, while using 13 TeV LHC data 
the observed signal strength is $\mu^{4\ell}_{gg{\rm F}} = 1.20^{+0.22}_{-0.21}$ \cite{Sirunyan:2017exp}.    
The current data in $H \to 4\ell$ channel alone cannot be used to constrain the parameters 
$c_{ZZ}$ and $c_{Z\square}$. Therefore, at present, any meaningful bounds on these parameters 
can be obtained by including data in other decay channels which have been observed in production modes
sensitive to $c_{ZZ}$ and $c_{Z\square}$, \textit{i.e.} 
vector boson fusion (VBF) and associate production of
Higgs and vector boson (VH)~\cite{Falkowski:2015fla}.

\subsubsection{$\chi^2$ fit with normalized distributions and asymmetries}
\label{susub:nor}
In the first stage of the analysis we consider normalized
distributions, and we look for the kinematic
observables turning out to be particularly sensitive to $c_{ZZ}$ and
$c_{Z\square}$ effects. The analysis is performed through a sample of  
$pp \to H \to 4\ell$ pseudo-events. 
For the sake of simplicity the sample is restricted to 
the ggF production mode.

The sample has been
generated by interfacing {\tt POWHEG}~\cite{Frixione:2007vw} 
to {\tt Hto4l}, according to the procedure described in
Ref.~\cite{Boselli:2015aha} and exploiting the Narrow Width
Approximation (NWA).   
The expected number of SM events is derived by assuming 3
$\mathrm{ab}^{-1}$ of integrated luminosity. The adopted values for 
ggF cross section and $H \to 2\ell2\ell'$ $\(\ell,\ell'=e,\mu\)$ 
branching ratios are 
taken from Ref.~\cite{deFlorian:2016spz}. The events are then selected   
according to the experimental cuts adopted in
ATLAS~\cite{Aad:2014tca}. Eventually,  
the accepted events are scaled down by 20\% to take into account 
the lepton reconstruction efficiency (95\% for each lepton), leading 
to a sample of $~\sim 6000$ reconstructed events, in good agreement 
with the number found in Ref.~\cite{Anderson:2013afp}. Besides the
distributions defined in the previous section, the two asymmetries 

\begin{eqnarray}
  \mathcal{A}^{\(3\)}_\phi &=& \frac{1}{\sigma}\int \rmd\Omega \,
  \textrm{sgn}\lg \cos \phi \rg \frac{\rmd \sigma}{\rmd
    \Omega}, \label{eqn:asy1} \\
  \mathcal{A}_{c \theta_1 c \theta_2} &=& \frac{1}{\sigma}\int \rmd\Omega \,
  \textrm{sgn}\lg \cos\theta_1 \cos\theta_2 \rg \frac{\rmd
    \sigma}{\rmd \Omega}, \label{eqn:asy2}
\end{eqnarray}
are sensitive to CP-even $D=6$ coefficients 
(as already pointed out in Ref.~\cite{Beneke:2014sba}). 
In the above definitions $\phi$ is the angle between the decay planes of the
two intermediate vector bosons, while $\theta_1$ ($\theta_2$) is the angle between
the lepton produced in the decay of the non-resonant (resonant) $Z$ boson
and the direction opposite the Higgs boson, in the non-resonant (resonant) $Z$ boson rest-frame.

The $\chi^2$ for distributions and asymmetries can be written as follows: 

\begin{eqnarray}
  \chi^2_{\mathrm{D}} &=&  \sum_{i=1}^{N_{D}}
  \frac{\(f_{i}^{\mathrm{BSM}} -
    f_{i}^{\mathrm{SM}}\)^2}{\sigma_{i}^2},
  \label{eqn:chid} \\
  \chi^2_{\mathcal{A}} &= &
  \frac{\(\mathcal{A}^{\mathrm{BSM}} -
    \mathcal{A}^{\mathrm{SM}}\)^2}{\sigma_{\mathcal{A}}^2}.
  \label{eqn:chias}
\end{eqnarray}
$N_D$ is the number of bins of the $D$-th distribution.   
The quantity $f_{i}^{\mathrm{SM}}$ is the fraction of events,
generated as described above, falling in the  $i$-th bin of the
SM distribution, while  
$f_{i}^{\mathrm{BSM}}$ is the fraction of expected events in the presence
of a given combination of $c_{ZZ}$ and $c_{Z\square}$. This last
quantity is calculated by 
reweighting the events with a program in which the {\tt Hto4l} BSM
matrix elements have been implemented. As we deal with normalized
quantities, we assume that the systematic and theoretical
uncertainties are cancelled to a large extent in the ratio. 
Accordingly, $\sigma_{i}$ and $\sigma_{\mathcal{A}}$ 
in Eqs.~(\ref{eqn:chid}- \ref{eqn:chid}) are just the one-sigma
statistical uncertainties
\begin{eqnarray}
  \sigma_{i} &\approx& \frac{\sqrt{n_i}}{N}, \label{eqn:sig-d} \\
  \sigma_{\mathcal{A}} &=& \sqrt{\frac{1-\mathcal{A}^2}{N}} \approx
  \frac{1}{\sqrt{N}}, \label{eqn:sig-a}  
\end{eqnarray}
where $n_i$ is the number of events falling in the $i$-th bin. 
\begin{figure}[t]
 \begin{subfigure}{0.5\linewidth}
   \includegraphics[width=8cm,clip]{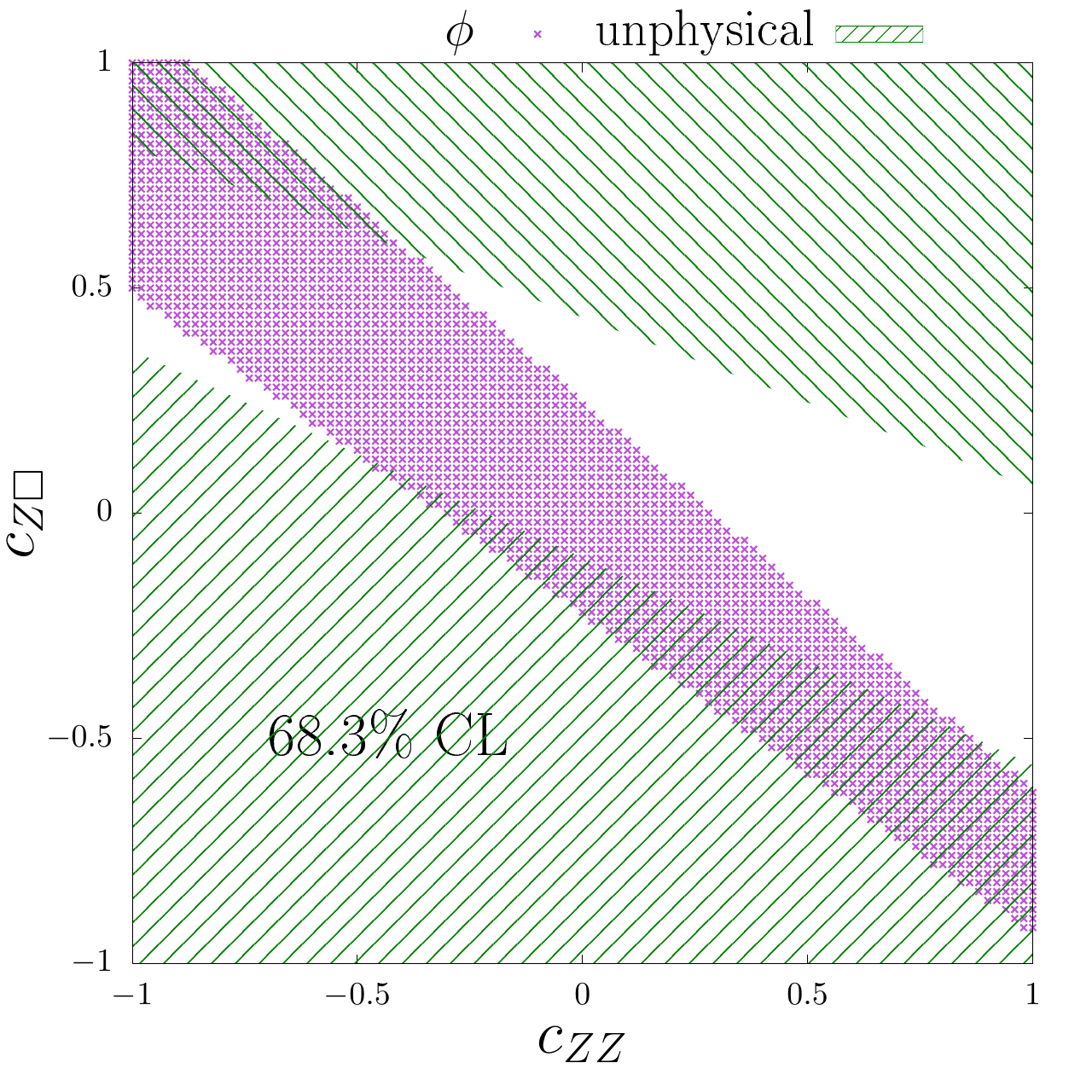}
     \end{subfigure}
    \begin{subfigure}{0.5\linewidth}
   \includegraphics[width=8cm,clip]{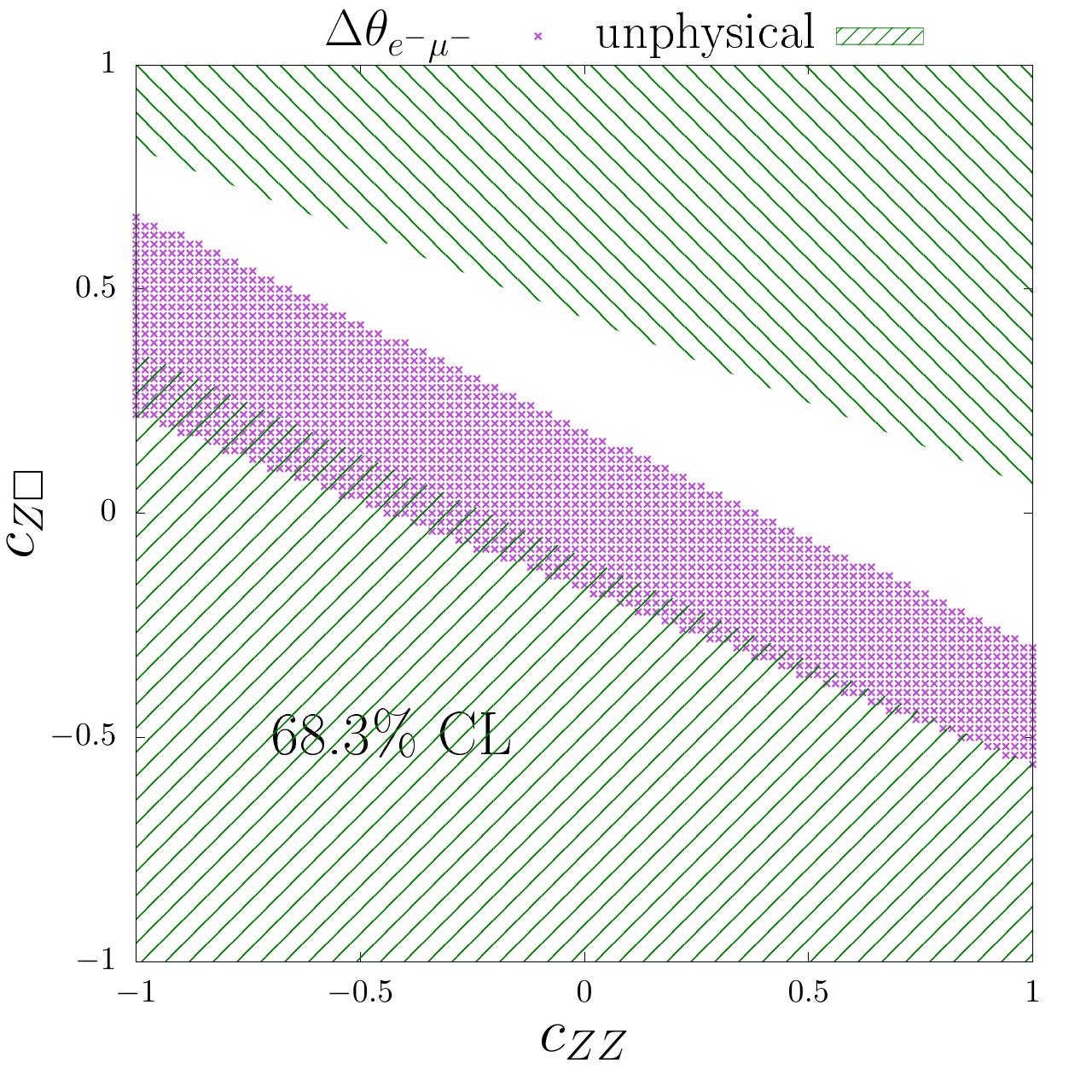}
     \end{subfigure}
    \begin{subfigure}{0.5\linewidth}
   \includegraphics[width=8cm,clip]{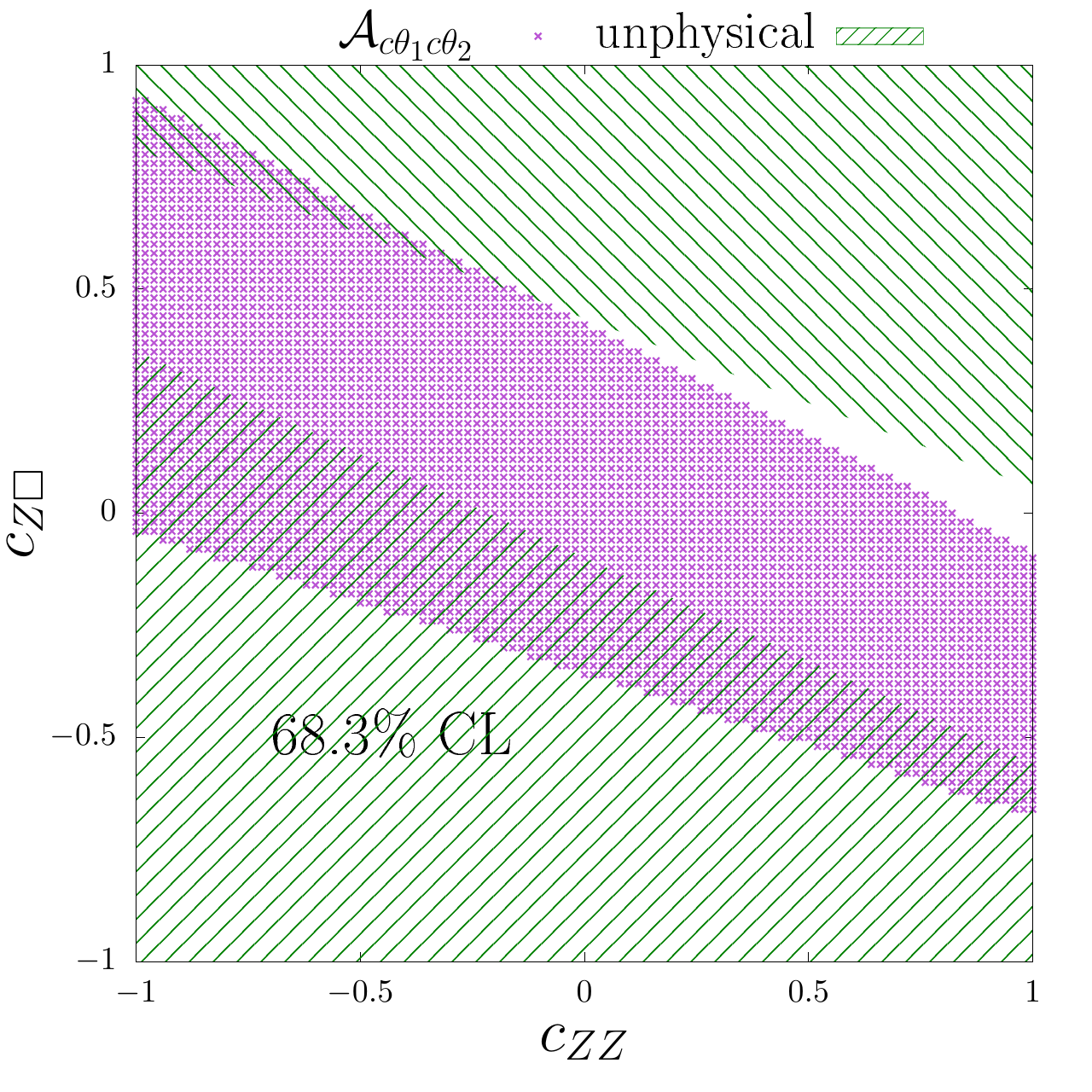}   
     \end{subfigure}
    \begin{subfigure}{0.5\linewidth}
   \includegraphics[width=8cm,clip]{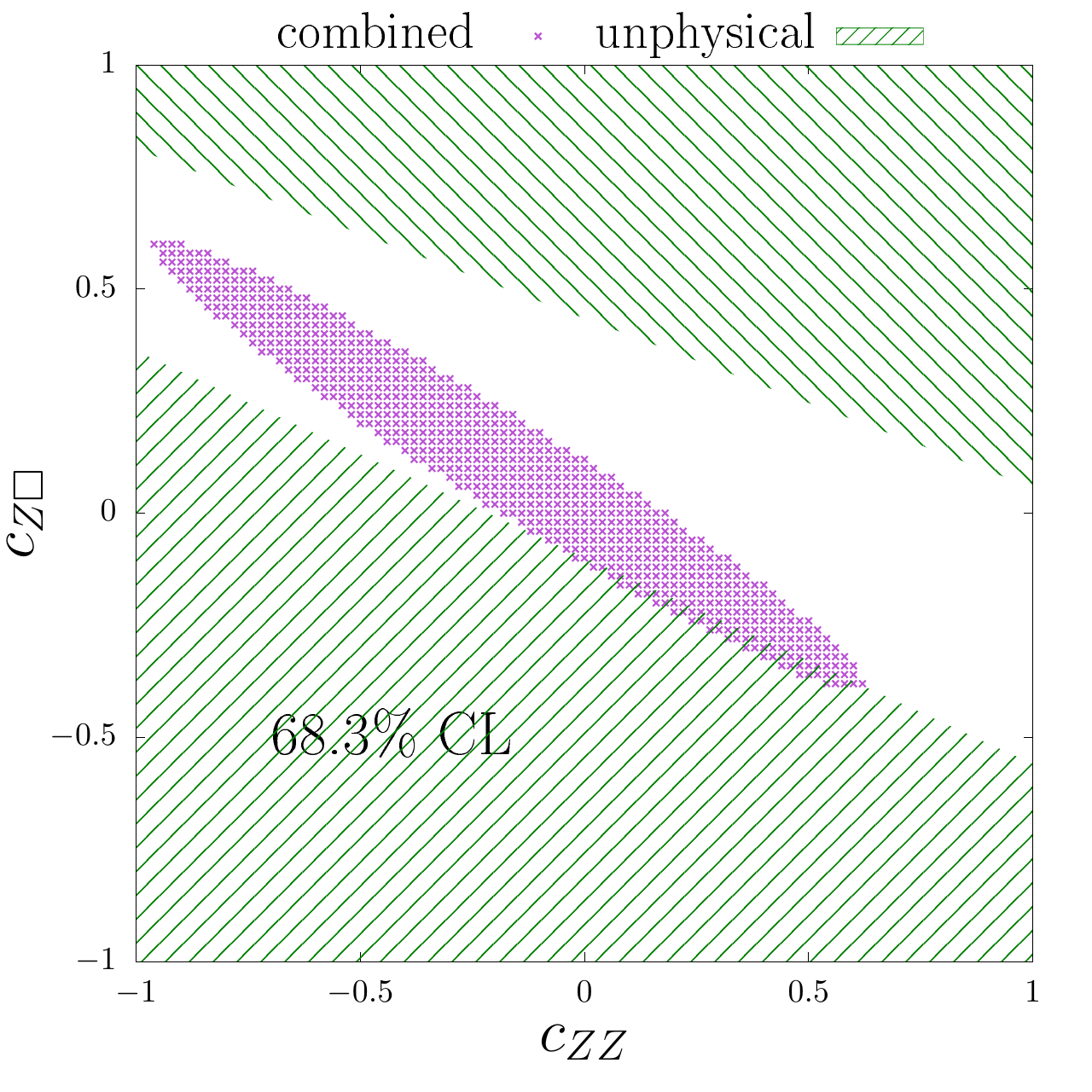}  
     \end{subfigure}
\caption{\footnotesize{Contour plot at 68.3\% CL for the angles $\phi$,
$\Delta\theta_{e^-\mu^-}$ (upper panels) and $\mathcal{A}_{c \theta_1 c \theta_2}$
    (lower-left panel). In the lower-right panel we display the
    contour plot derived from the $\chi^2_{\mathrm{comb.}}$ (Eq.~(\ref{eqn:chitot}). 
    The contour plots are derived according to the procedure detailed in Sec.~\ref{susub:nor}. 
    The green regions 
    correspond to values of the effective coefficients leading one of the quantities in 
    Eqs.~(\ref{eqn:xsvbf}-\ref{eqn:xszh}) to negative values.}}
 \label{fig:differential}
 \end{figure}
The 68.3\% Confidence Level (CL) contour plots for the aforementioned
distributions and asymmetries are displayed in
Fig.~\ref{fig:differential}. The contour plot for the asymmetry
$\mathcal{A}^{(3)}_\phi$ overlaps exactly the one for the
$\phi$ angle and therefore is not shown. 
The contour plot for the combined $\chi^2$ defined by the sum

\begin{equation}
\chi^2_{\mathrm{comb.}} = \chi^2_{D} +\chi^2_{\mathcal{A}}
\label{eqn:chitot} 
\end{equation} 
is also displayed. In the next section, we perform a global analysis using signal 
strengths where production channels other than ggF are also considered. Unlike ggF, 
the production channels VBF and $VH$ depend upon $c_{ZZ}$ and $c_{Z\square}$ and this 
dependence is quite strong (\ref{eqn:xsvbf}-\ref{eqn:xszh}). This feature is taken into 
account in Fig.~\ref{fig:differential}. The regions marked by green lines correspond to 
parameter-space points, \textit{i.e.} to $c_{ZZ}$ and $c_{Z\square}$ values, driving any 
of these cross sections to negative values. We remark that these ``unphysical'' regions 
arise because in the linear approximation, the cross sections are not positive definite.
  
Few remarks are in order:

\begin{itemize}
\item the $\chi^2$ analysis of single distributions and asymmetries
  is not sufficient to get closed contour plots in the range (-1,1) for $c_{ZZ}$ 
  and $c_{Z\square}$. This is mainly motivated by the fact
  that one can choose values for $c_{ZZ}$ and $c_{Z\square}$ whose
  effects cancel in the sum;
  \item among the four angular observables taken into account, the 
  asymmetry $\mathcal{A}_{c \theta_1 c  \theta_2}$ turns out to be
  the least sensitive observable to $c_{ZZ}$ and $c_{Z\square}$;
\item the negative correlation between $c_{ZZ}$ and $c_{Z\square}$
  resulting from the analysis of Run-I data arise in our analysis of
  distributions as well. However, the correlation for $\phi$
  distribution and $\mathcal{A}^{(3)}_\phi$ asymmetry is larger
  than for $\Delta\theta_{e^-\mu^-}$ and
  $\mathcal{A}_{c \theta_1 c \theta_2}$. This feature allows to rule 
  out a region of the parameter space, with $c_{ZZ} > 0$ and
  $c_{Z\square} < 0$. 
  \item the contour plot derived from Eq.~(\ref{eqn:chitot}) lies inside 
  the ``physical region'' of the parameter space. The constraints on 
  $c_{ZZ}$ and $c_{Z\square}$ are tighter, while the correlation is not   
  removed.
\end{itemize}



 \subsubsection{$\chi^2$ fit using signal strengths} 
   
At the high-luminosity run of the LHC, the $H \to 4\ell$ decay channel is likely to be observable in production channels other than ggF. 
Moreover, we will also have access to the kinematic distributions in $H \to 4\ell$. Therefore, in the context of HL-LHC, it would be interesting to study the sensitivity of the future data in constraining parameters $c_{ZZ}$ and $c_{Z\square}$ mainly using the $H \to 4\ell$ decay channel. 
Our main motivation is to highlight the effect of angular distributions $\phi$ and $\Delta\theta_{e^-\mu^-}$ in the fit.

   
   Our analysis is based on minimizing the $\chi^2$ function built using signal strengths as observables. The signal strength in a given production channel $i$ and decay channel $f$
   is defined as $\mu_i^f = \mu_i \times \mu^f$, where
   \begin{eqnarray}
    \mu_i = \frac{\sigma_i^{\rm BSM}}{\sigma_i^{\rm SM}},\qquad \mu^f = \frac{{\rm BR}_f^{\rm BSM}}{{\rm BR}_f^{\rm SM}}.
   \end{eqnarray}
   The expressions for $\mu_i$ and $\mu^f$, in presence of parameters $c_{ZZ}$ and $c_{Z\square}$ are 
   taken from~\cite{DiVita:2017eyz} where linear approximation in parameters is adopted. The $\mu^{4\ell}$ both at 
   inclusive and differential levels is calculated using the {\tt Hto4l} code with input parameter 
   choice and kinematic cuts mentioned above. We have summarized all these expressions in Appendix~\ref{appen:signal-strengths}.
   The $\chi^2$ function in terms of signal strengths is given by,
   \begin{equation}
    \chi^2 = \sum_{i,f} \frac{(\mu_i^{f, \rm exp} - \mu_i^f)^2}{(\sigma_i^f)^2},
   \end{equation}
   where the one-sigma uncertainties $\sigma_i^f$ are taken from Ref.~\cite{ATL-PHYS-PUB-2014-016}. For future data we assume $\mu_i^{f, \rm exp} = 1.0$ and $\sigma_i^f$ 
   is the expected uncertainty in a given 
   channel $\{i,j\}$ at the inclusive level which includes theory, experimental systematic 
   and statistical uncertainties. When using the kinematic distribution in the fit we keep 
   theory and experimental systematic uncertainties the same in all the bins, while the statistical 
   uncertainty in each bin is scaled up by the fraction of events falling in it. Following the 
   nature of event distributions in Figs.~\ref{fig:phi1} and ~\ref{fig:cos}, we divide $\phi$ distribution in 3 bins 
   ($[0,90]; [90,270]; [270,360]$) and $\Delta\theta_{e^-\mu^-}$ distribution in two bins ($[0,100]; [100,180]$). 
   We find that the effect of $\Delta\theta_{e^-\mu^-}$ distribution in the fit is similar but less
   important than that of $\phi$ distribution. All the results in the following are presented using 
   $\phi$ distribution.

 \begin{figure}[t]
 \begin{subfigure}{0.5\linewidth}
 \includegraphics[width=8cm,clip]{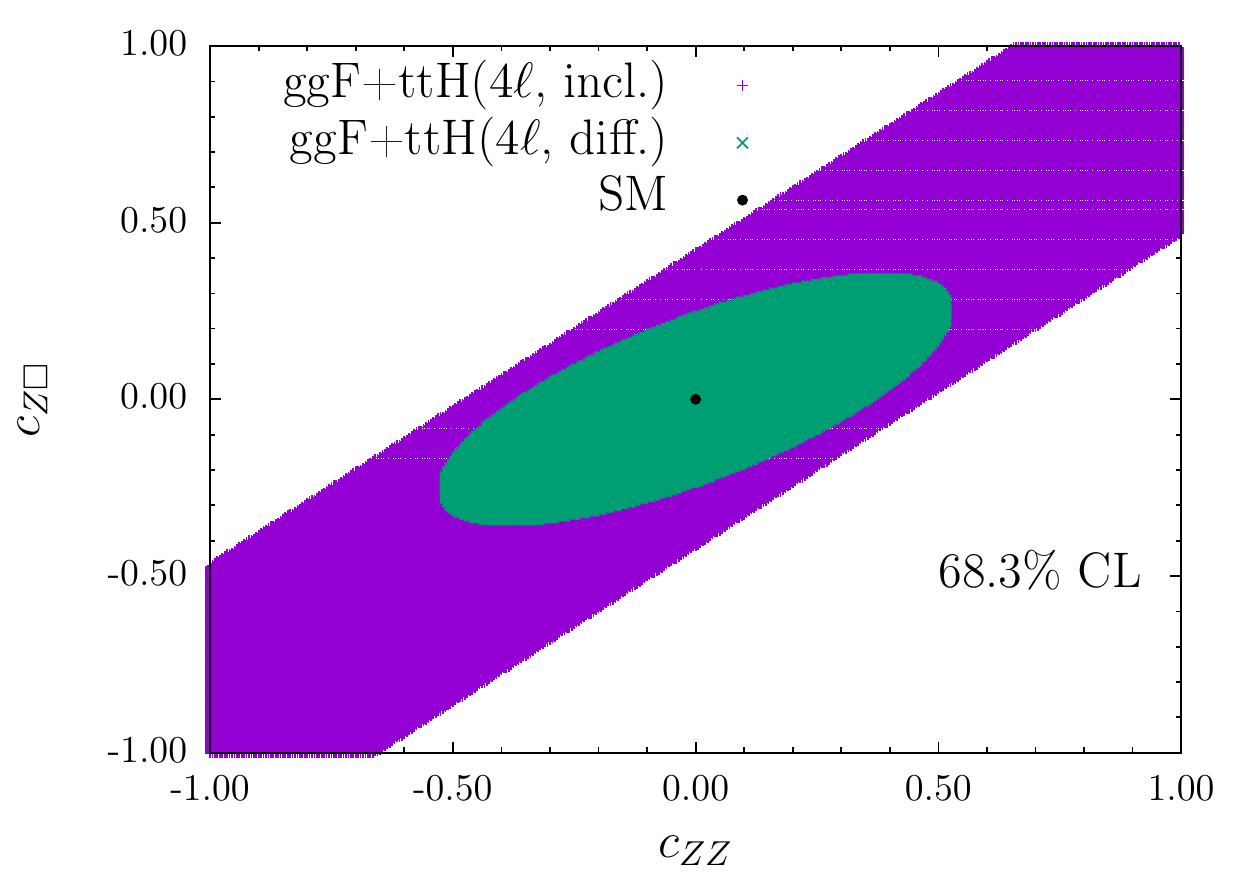}\subcaption{}\label{fig:fit-4l-a}  
 \end{subfigure}
  \begin{subfigure}{0.5\linewidth}
 \includegraphics[width=8cm,clip]{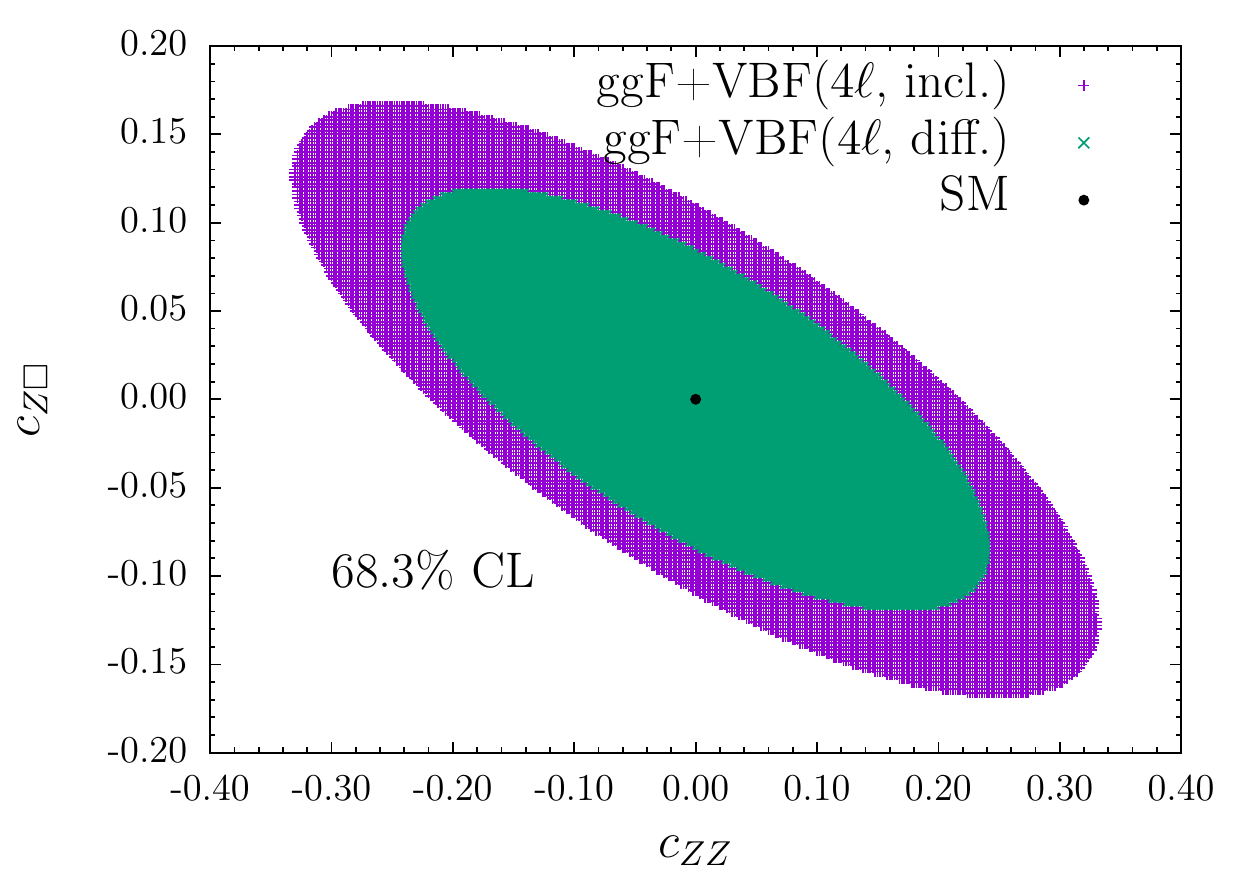}\subcaption{}\label{fig:fit-4l-b}  
  \end{subfigure}
  \begin{subfigure}{0.5\linewidth}
 \includegraphics[width=8cm,clip]{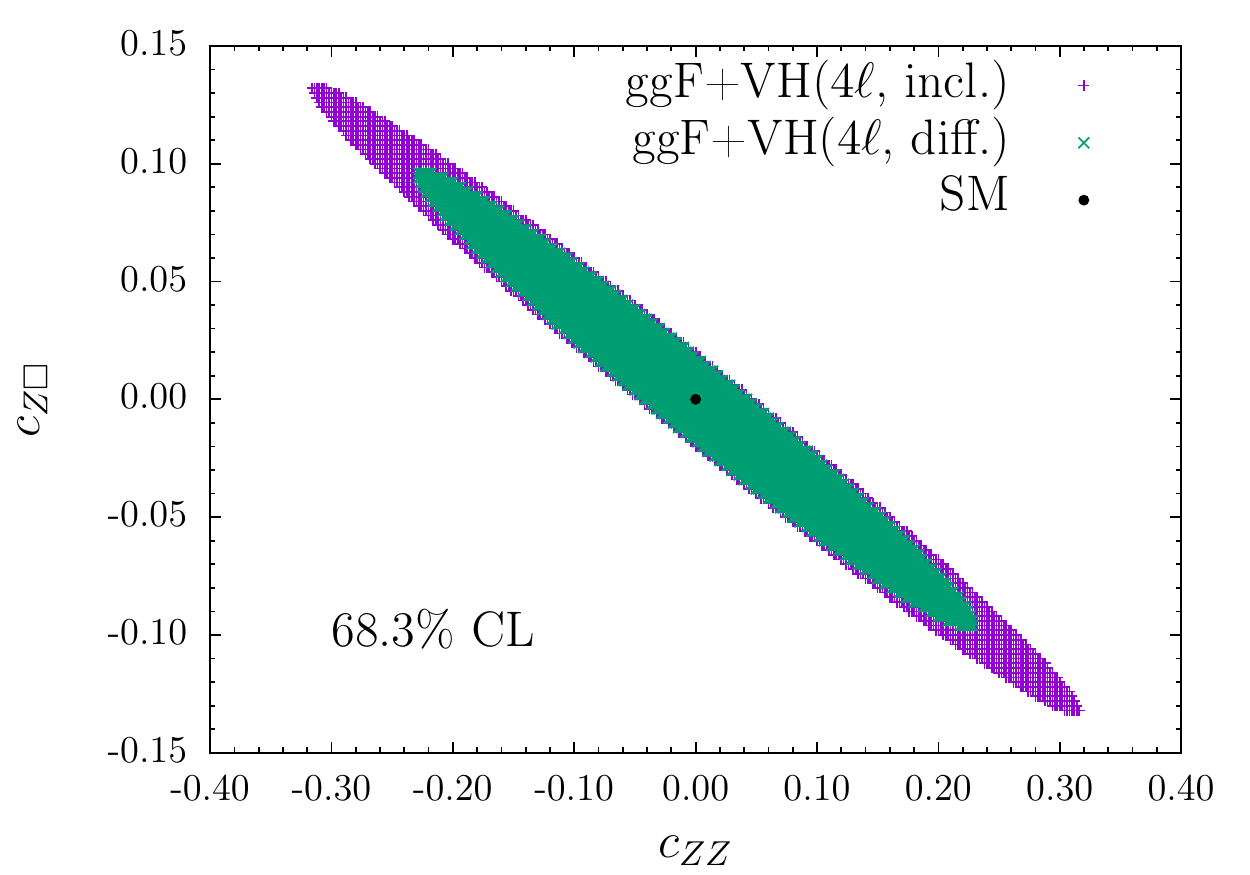}\subcaption{}\label{fig:fit-4l-c} 
  \end{subfigure}
  \begin{subfigure}{0.5\linewidth}
 \includegraphics[width=8cm,clip]{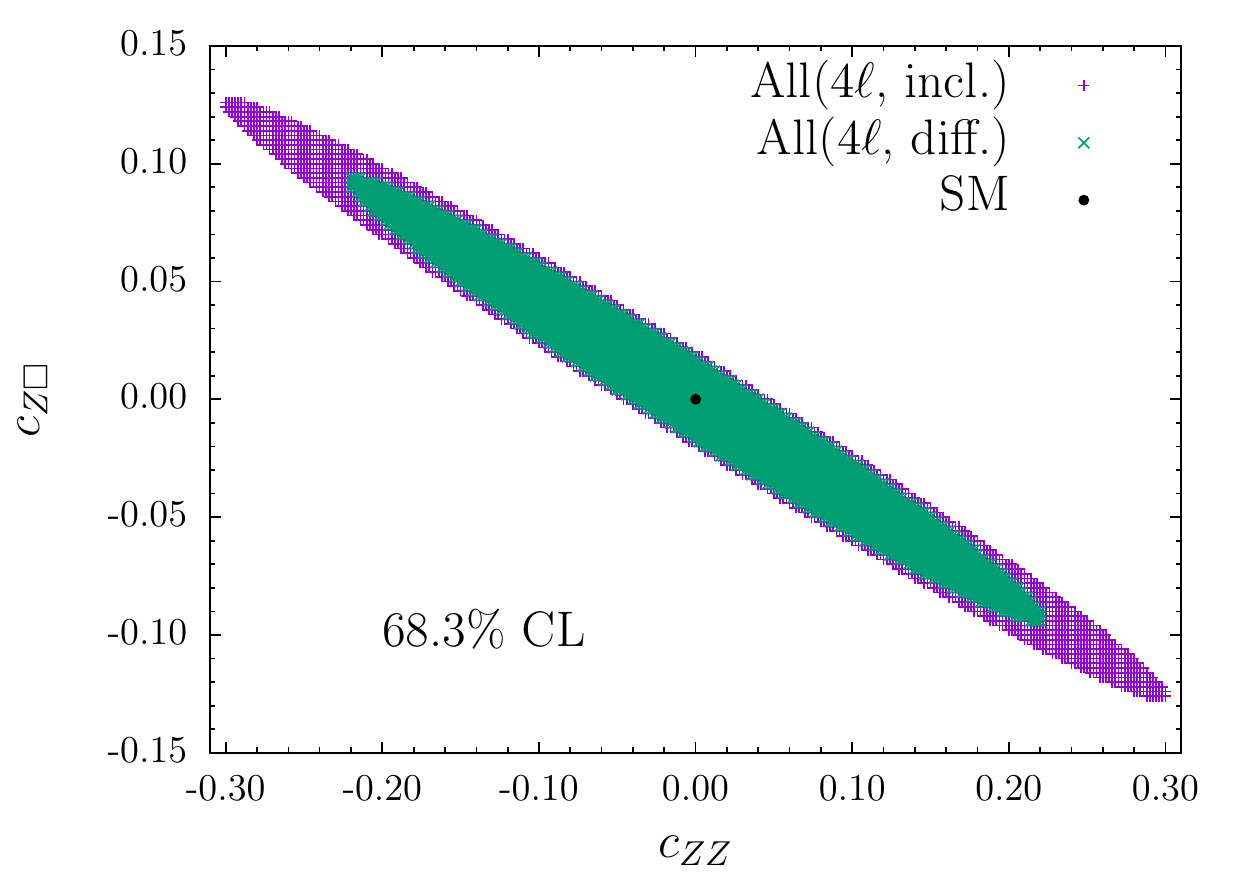}\subcaption{}\label{fig:fit-4l-d} 
  \end{subfigure}
 \caption{\footnotesize{A comparison of global fits using inclusive (purple) and differential (green) 
 information in the definitions of signal strength for the $H\to 4\ell$ decay. Various 
 cases based on the choice of production channels included in the fit are displayed. 
 By definition, the best fit for the parameters corresponds to the SM prediction {\it i.e.} (0,0). }}\label{fig:fit-4l}
 \end{figure}
  
  In Fig.~\ref{fig:fit-4l} we display the region plots with 68.3\% CL in $c_{ZZ}-c_{Z\square}$ plane. 
  In each plot we compare the fits obtained by using the decay signal strength $\mu^{4\ell}$ at the 
  inclusive (1 bin) and differential (3 bins) levels. 
  To understand the 
  effect better we divide the fit in many categories depending upon which production channels are used in the 
  fit. The differential effects are the largest when only ggF and $ttH$ production channels are 
  included in the fit (Fig.~\ref{fig:fit-4l-a}). This is not surprising given the fact that both these production channels 
  do not depend on parameters $c_{ZZ}$ and $c_{Z\square}$. The positive correlation between 
  the parameters is governed by the decay signal strength $\mu^{4\ell}$ (see Eqs.~(\ref{eqn:4l-incl}-\ref{eqn:4l-bin3})). 
  When we use ggF and VBF channels (Fig.~\ref{fig:fit-4l-b}), the correlation becomes negative due to a stronger dependence 
  of $\mu_{\rm VBF}$ (\ref{eqn:xsvbf}) on the parameters which is opposite in nature to that of $\mu^{4\ell}$. 
  Once again, the $\phi$ distribution in the fit improves the bounds significantly. Using ggF and $VH$ in the fit (Fig.~\ref{fig:fit-4l-c})
  the constraints on the parameters become tighter and the effect of including distributions is mostly 
  visible at the edges. 
  Note that the constraints become stronger because the dependence of $VH$ channel on 
  parameters is very strong (\ref{eqn:xswh},~\ref{eqn:xszh}). Thus, when all the production channels are combined (Fig.~\ref{fig:fit-4l-d}) the 
  constraints on parameters are governed by the production channels rather than the 
  $H \to 4\ell$ decay channel and the distribution still leads to a noticeable 
  improvement in the fit. 
  In this case, we also derive $1\sigma$ constraints on each parameter when the 
  other parameter is ignored in the fit. The $1\sigma$ errors on $c_{ZZ}$ and $c_{Z\square}$ resulting from the inclusive (differential) 
  fit are $\pm 0.032~(\pm 0.026)$ and  $\pm 0.014~(\pm 0.013)$ respectively.

 \begin{figure}[t]
  \begin{subfigure}{0.5\linewidth}
 \includegraphics[width=8cm,clip]{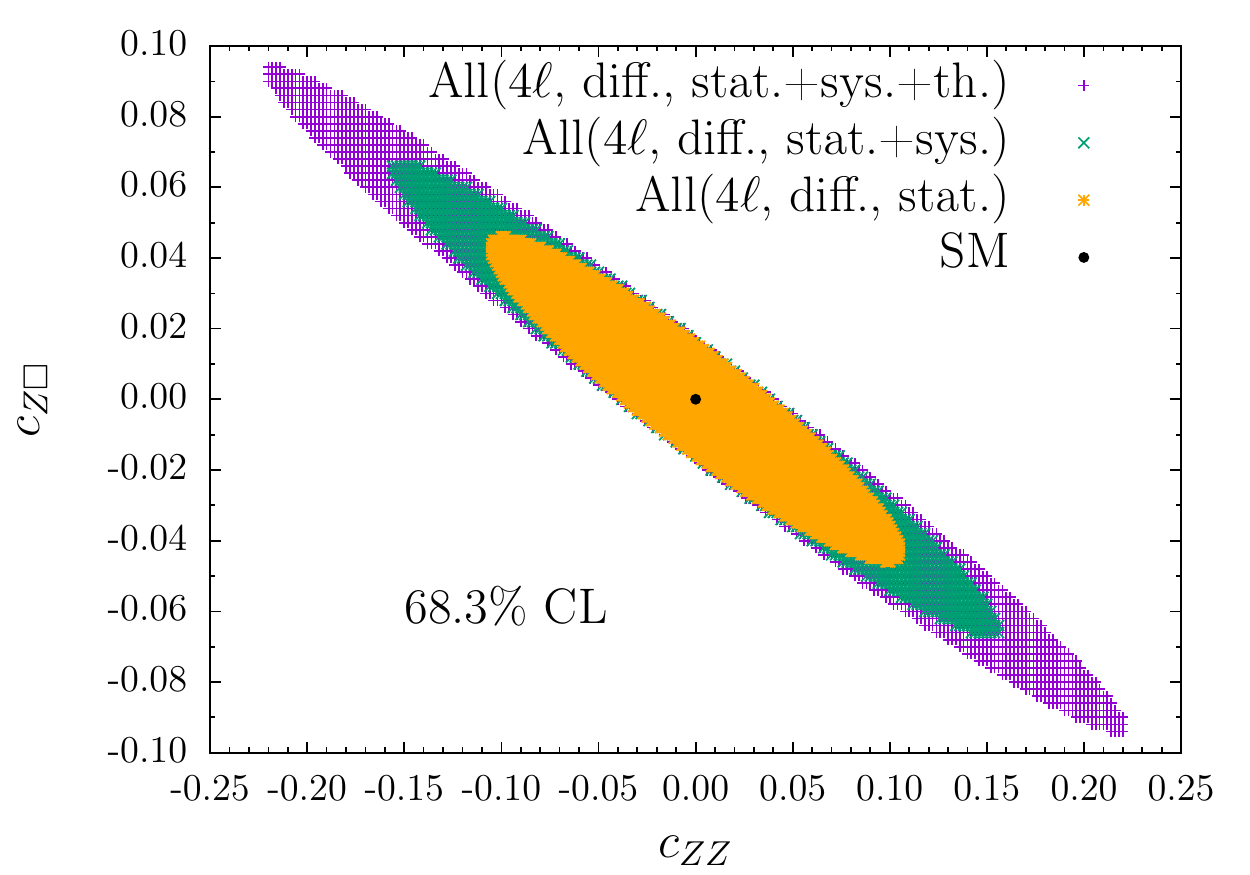}\subcaption{}\label{fig:fit-all-a}
   \end{subfigure}
  \begin{subfigure}{0.5\linewidth}
 \includegraphics[width=8cm,clip]{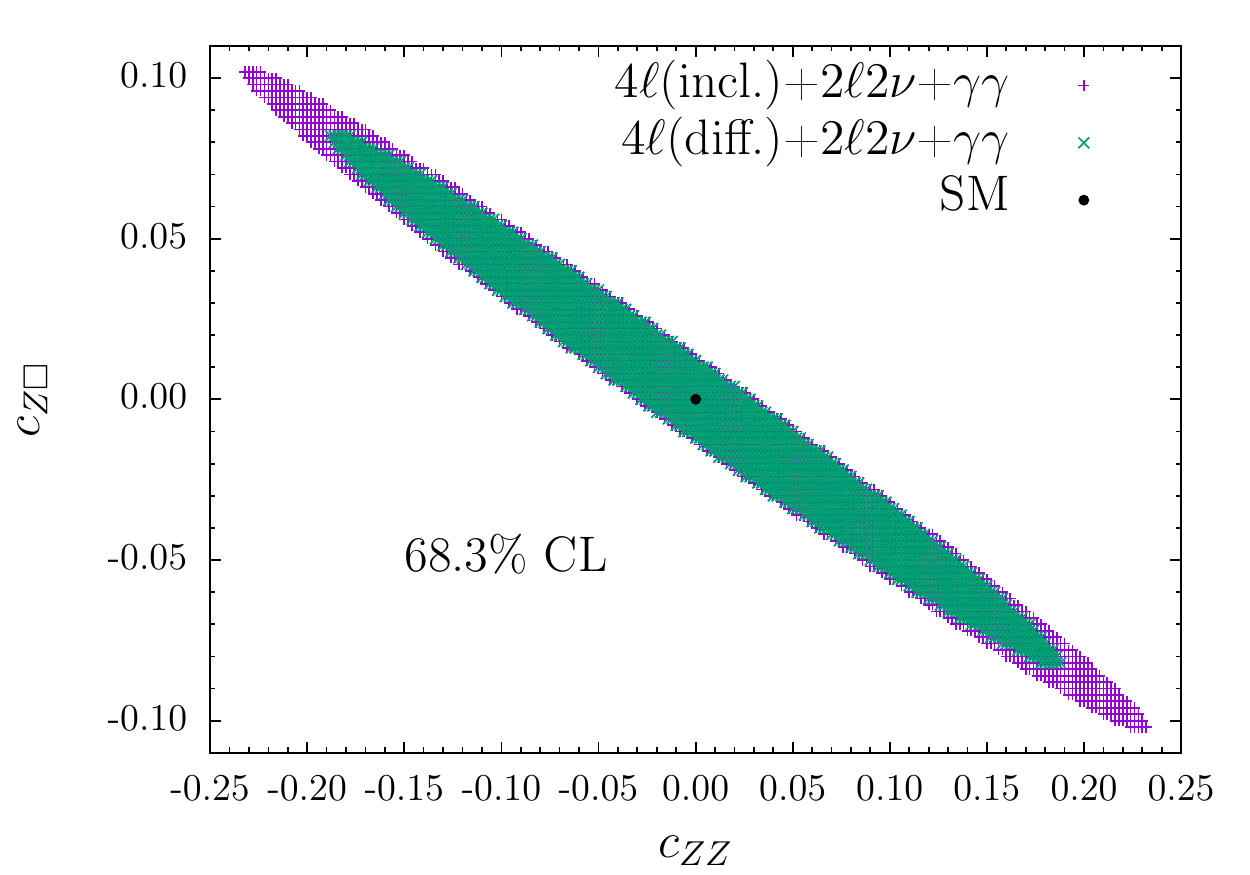}\subcaption{}\label{fig:fit-all-b}
   \end{subfigure}
  \caption{\footnotesize{(a) Effect of various uncertainties on the fit when differential information 
  in the $4\ell$ channel is used. (b) A comparison of inclusive (purple) and differential (green) 
  information for $H \to 4\ell$ in the fit when data in other decay channels are also 
  included. In each decay mode, all the relevant production channels are taken into account. }}\label{fig:fit-all}
 \end{figure}

  In Fig.~\ref{fig:fit-all-a}, we relax theoretical and experimental systematic uncertainties 
  in the future data, and perform the fit using only $\mu^{4\ell}$ 
  at the differential level in all production channels. Clearly, improvements in precision calculations would 
  reduce theoretical uncertainties, allowing to put tighter constraints on the parameters. 
  We also study the relevance of $\phi$ distribution in the fit when other decay channels like 
  $2\ell2\nu$ and $\gamma\gamma$, which depend on the parameters $c_{ZZ}$ and $c_{Z\square}$,
  are included (Fig.~\ref{fig:fit-all-b}). The $\gamma\gamma$ decay channel is included in all the production channels 
  while, the $2\ell2\nu$ decay channel is included only in ggF and VBF production channels. 
  Notice that the $\gamma\gamma$ partial decay width does not depend on the parameters at the 
  LO and in our analysis we have used the one-loop expression derived in ~\cite{Hartmann:2015aia} 
  and quoted in ~\cite{DiVita:2017eyz} for the Higgs basis parameters. 
  At inclusive level, the $1\sigma$ errors on $c_{ZZ}$ and $c_{Z\square}$ become $\pm 0.02$ and $\pm 0.01$ respectively.
  These errors on individual parameters are consistent with those obtained in Ref.~\cite{DiVita:2017eyz}.
  We find that the improvements in bounds due to distribution are marginal but still visible.



\section{Conclusions}

In the present work we have investigated possible NP effects in the
Higgs decay into four charged leptons using an EFT approach to
Higgs interactions. We have adopted the Higgs basis for the
computation of the BSM matrix elements for the $H \to 2e2\mu$ and
$H \to 4e/4\mu$ channels. We considered both CP-even and CP-odd
operators and we mostly focused on those parameters,  
which are weakly constrained by LHC Run-I data.
Since the $H \to 4\ell$ 
channel can provide information about the presence of $ZH\ell\ell$
contact interactions, we have also  
considered the scenario in which they are
independent of anomalous  $Z\ell\ell$ interactions and therefore unconstrained by electroweak
precision data.

For the sake of illustration, we have presented numerical results
for the $H\to 2e2\mu$ channel. As a first step, we have analyzed the impact of
$D=6$ operators on the partial decay width. As the information on partial decay width is not sufficient in 
discriminating different parameters, we have also studied some kinematic 
distributions of particular experimental interest. We have found that with
the help of  
angular observables ($\phi$ and $\Delta\theta_{e^-\mu^-}$) it is possible 
to distinguish different parameters with values which would lead to
the same modification 
in the partial decay  width. In the case of ${\tilde c}_{Z\gamma}$,
the angle $\phi$  
would be quite useful in deriving a stronger constraint on the
parameter as it captures  
the information on the linear piece in the parameter. 

{ Aiming to assess the impact of differential information in future analyses, we have performed a global 
analyses in the context of HL-LHC. 
From the preliminary study on differential distributions and asymmetries we derived that the angle $\phi$ 
is the most sensitive observable to $c_{ZZ}$ and $c_{Z\square}$ effects. 
In our global analysis based on signal strengths we find that the impact of the angular information is quite 
dependent on the production channels that are taken into account. The largest improvements are observed in the 
ggF and $ttH$ channels. When also VBF and $VH$ channels are included, the benefits coming from the inclusion of 
angular information are moderate but are still noticeable. More sophisticated analyses, where other coefficients 
and differential information coming from other production channels are also considered, are beyond the purposes 
of this work and will be considered in future extensions of the present study.} 

The above phenomenological study has been carried out through a new
version of the {\tt Hto4l} event generator, which allows to study  the
effects of $D=6$ operators in the $H \to 2e2\mu$ and
$H \to 4e/4\mu$ channels. The BSM matrix elements are 
calculated in the Higgs basis.
The code also allows independent calculations in SILH and Warsaw bases.
As an
option, the possibility of including quadratic $D=6$ contributions (of
the order of $1/\Lambda^4$) on top of pure $1/\Lambda^2$ interference
contributions is given.
Since it  can be easily interfaced with
any event generator for the Higgs production, {\tt Hto4l} can be used in
association with other MC tools for the full simulation of Higgs
events in an EFT framework.


\section*{Acknowledgments}
We gratefully acknowledge A. Pomarol and F. Riva for useful
discussion. We would also like to thank Ken Mimasu and Michael 
Trott for discussions  
on the implementation of independent dictionaries for 
SILH and Warsaw bases. AS would like to
acknowledge many fruitful discussions on EFT he had  
with speakers and the participants of the CP3 weekly Pheno meeting. 
AS is supported by the MOVE-IN Louvain Cofund grant and the IISN
``Fundamental interactions'' convention 4.4517.08.
\appendix

\newpage

\section{$H \to 4e$  partial decay width }\label{appen:H4e}
The coefficients in Eq.~(\ref{eqn:GGG}) for the decay channel $H \to 4e$   
are given by 

\begin{equation*}
  X'_i = 
  \begin{pmatrix}
    2.00 & -0.0108 & 0.169 & -0.228 & 0.261 & -8.83 & 6.97 & 4.71 & -3.66
  \end{pmatrix},
\end{equation*}
\begin{equation*}
  X'_{ij} =
  \begin{pmatrix}
    1.00 & 0.0108 & 0.169 & -0.228 & 0.261 & -8.83 & 6.97 & 4.71 & -3.66 \\
    0 & 0.0801 & 0.0732 & -0.00351 & -0.00516 & -0.154 & -0.289 &
    0.127 & 0.264 \\ 
    0 & 0 & 0.964 & -0.514 & -0.747 & -3.47 & -2.81 & 2.98 & 2.88 \\
    0 & 0 & 0 & 0.125 & 0.317 & 2.34 & 0.872 & -1.70 & -1.00 \\
    0 & 0 & 0 & 0 & 0.290 & 0.944 & 3.52 & -1.26 & -2.66 \\
    0 & 0 & 0 & 0 & 0 & 25.9 & -14.8 & -27.9 & 2.56 \\
    0 & 0 & 0 & 0 & 0 & 0 & 22.1 & 5.56 & -25.5 \\
    0 & 0 & 0 & 0 & 0 & 0 & 0 & 8.82 & -1.44 \\
    0 & 0 & 0 & 0 & 0 & 0 & 0 & 0 & 8.06
  \end{pmatrix}, \\
\end{equation*}
\begin{equation*}
  {\tilde X'}_{ij} = 
  \begin{pmatrix}
    0.0701 & 0.116 & -0.00181 \\
    0 & 0.373 & 0.119 \\
    0 & 0 & 0.00198
  \end{pmatrix}.
\end{equation*}

\section{EFT Dictionaries}\label{appen:dictionary} 
In the latest version of the {\tt  Hto4l} code, the calculation of $H \to 4\ell$ 
BSM matrix elements can be performed {\it independently} in the Higgs basis, 
the SILH basis and the Warsaw basis. The BSM matrix elements are implemented 
in the Higgs basis. For predictions in SILH and Warsaw bases the Higgs basis 
parameters are seen just as the coefficients corresponding to specific Lorentz 
structures in the Feynman rules of section~\ref{sec:calculation}, and following 
dictionaries between the Higgs basis parameters and 
the Wilson coefficients of SILH and Warsaw bases are used.

\subsection{SILH}
 \begin{eqnarray}
  c_{\gamma\gamma} &=& \frac{16}{g_2^2} K_\gamma \\
  {\tilde c}_{\gamma\gamma} &=& \frac{16}{g_2^2} {\tilde K}_\gamma \\
  c_{Z\gamma} &=& -\frac{2}{g_2^2} \Big( K_{HW} - K_{HB} + 8 \frac{g_1^2}{g_1^2+g_2^2} K_\gamma \Big)  \\
  {\tilde c}_{Z\gamma} &=& -\frac{2}{g_2^2} \Big( {\tilde K}_{HW} - {\tilde K}_{HB} + 8 \frac{g_1^2}{g_1^2+g_2^2} {\tilde K}_\gamma \Big)  \\
  c_{\gamma\square} &=& \frac{2}{g_2^2} \Big( K_W -K_B + K_{HW} - K_{HB} \Big) \\
  \delta c_Z &=&  2 K_T -\frac{1}{2} K_H + \frac{2g_1^2}{g_1^2+g_2^2} (K_W+K_B)\\
  c_{ZZ} &=&   - \frac{4}{g_1^2+g_2^2} \Big( K_{HW} + \frac{g_1^2}{g_2^2}( K_{HB} - 4 \frac{g_1^2}{g_1^2+g_2^2} K_\gamma) \Big) \\
  {\tilde c}_{ZZ} &=&   - \frac{4}{g_1^2+g_2^2} \Big( {\tilde K}_{HW} + \frac{g_1^2}{g_2^2}( {\tilde K}_{HB} - 4 \frac{g_1^2}{g_1^2+g_2^2} {\tilde K}_\gamma) \Big) \\
  c_{Z\square} &=&  \frac{2}{g_2^2} \Big( K_W + K_{HW} + \frac{g_1^2}{g_2^2} (K_B + K_{HB}) \Big) \\
%
  \delta g_L^{HZ\ell\ell} &=& -\frac{1}{2}\Big(K_{H\ell} + K'_{H\ell}\Big) \\ 
  \delta g_R^{HZ\ell\ell} &=& -\frac{1}{2} K_{He} \\
  \delta g_L^{Z\ell\ell} &=& -\frac{1}{2}\Big(K_{H\ell} + K'_{H\ell} \Big) + \frac{1}{2} \frac{g_1^2(g_2^2-g_1^2)}{(g_1^2+g_2^2)^2}  (K_W+K_B) \\ 
  \delta g_R^{Z\ell\ell} &=& -\frac{1}{2} K_{He} + \frac{g_1^2g_2^2}{(g_1^2+g_2^2)^2} (K_W+K_B).
 \end{eqnarray}
 
 The corrections to weak boson masses are given by, 
 \begin{eqnarray}
  \delta m_W^2 &=& 0, \\ 
  \delta m_Z^2 &=& \frac{1}{4} (g_1^2+g_2^2) v^2 \Big( -K_T +  \frac{2g_1^2}{g_1^2+g_2^2}  (K_W+K_B) \Big). 
 \end{eqnarray}

 The dependence of $s_W$ and $e$ on $g_1$ and $g_2$ becomes, 
 \begin{eqnarray}
 s_W^2 &=& \frac{g_1^2}{g_1^2+g_2^2} \Big( 1 -  \frac{g_1^2-g_2^2}{g_1^2+g_2^2}  (K_W+K_B) \Big) \\
 e &=& \frac{g_1 g_2}{\sqrt{g_1^2+g_2^2}} \Big( 1- \frac{g_1^2 }{g_1^2+g_2^2}  (K_W+K_B)  \Big).
\end{eqnarray}
 
\subsection{Warsaw}
 \begin{eqnarray}
  c_{\gamma\gamma} &=& 4 v^2 \Big(\frac{C_{HB}}{g_1^2} + \frac{C_{HW}}{g_2^2} - \frac{C_{HWB}}{g_1g_2}\Big) \\
  {\tilde c}_{\gamma\gamma} &=& 4 v^2 \Big(\frac{C_{H\tilde B}}{g_1^2} + \frac{C_{H\tilde W}}{g_2^2} - \frac{C_{H\tilde WB}}{g_1g_2}\Big) \\
  c_{Z\gamma} &=& 2 v^2 \frac{g_1^2+g_2^2}{g_1g_2}  \Big(\frac{2}{g_1g_2} (C_{HW}-C_{HB}) 
  - (\frac{1}{g_1^2}-\frac{1}{g_2^2})C_{HWB}\Big) \\
  {\tilde c}_{Z\gamma} &=& 2 v^2 \frac{g_1^2+g_2^2}{g_1g_2}  \Big(\frac{2}{g_1g_2} (C_{H\tilde W}-C_{H\tilde B}) 
  - (\frac{1}{g_1^2}-\frac{1}{g_2^2})C_{H\tilde WB}\Big) \\
  c_{\gamma\square} &=& 0 \\
  \delta c_Z &=& v^2 \Big( C_{H\square}+\frac{3}{4}C_{HD} + 2 \frac{g_1 g_2}{g_1^2+g_2^2} C_{HWB} \Big)  \\
  c_{ZZ} &=& 4v^2 \frac{1}{(g_1^2+g_2^2)^2} \Big( g_1^2 C_{HB} + g_2^2 C_{HW} + g_1g_2 C_{HWB} \Big) \\
  {\tilde c}_{ZZ} &=& 4v^2 \frac{1}{(g_1^2+g_2^2)^2} \Big( g_1^2 C_{H\tilde B} + g_2^2 C_{H\tilde W} + g_1g_2 C_{H\tilde WB} \Big) \\
  c_{Z\square} &=& 0 \\
%
  \delta g_L^{HZ\ell\ell} &=& -\frac{v^2}{2}\Big(C_{H\ell}^{(1)} + C_{H\ell}^{(3)}\Big) \\ 
  \delta g_R^{HZ\ell\ell} &=& -\frac{v^2}{2} C_{He} \\
  \delta g_L^{Z\ell\ell} &=& \frac{v^2}{2}\Big(-C_{H\ell}^{(1)} - C_{H\ell}^{(3)} + \frac{g_1g_2}{(g_1^2+g_2^2)^2} (g_2^2-g_1^2) C_{HWB} \Big) \\ 
  \delta g_R^{Z\ell\ell} &=& \frac{v^2}{2} \Big( -C_{He} + \frac{g_1g_2}{(g_1^2+g_2^2)^2} (2g_2^2) C_{HWB} \Big).
 \end{eqnarray}

 The corrections to weak boson masses are given by, 
 \begin{eqnarray}
  \delta m_W^2 &=& 0, \\
  \delta m_Z^2 &=& \frac{1}{4} (g_1^2+g_2^2) v^4 \Big(\frac{C_{HD}}{2} 
		+ 2 \frac{g_1 g_2}{g_1^2+g_2^2} C_{HWB} \Big).
 \end{eqnarray}
 
In addition, $s_W$ and $e$ have a modified dependence on $g_1$ and $g_2$ given by, 
\begin{eqnarray}
 s_W^2 &=& \frac{g_1^2}{g_1^2+g_2^2} \Big( 1 - \frac{g_2}{g_1} \frac{g_1^2-g_2^2}{g_1^2+g_2^2} v^2 C_{HWB} \Big) \\
 e &=& \frac{g_1 g_2}{\sqrt{g_1^2+g_2^2}} \Big( 1- \frac{g_1 g_2}{g_1^2+g_2^2} v^2 C_{HWB} \Big).
\end{eqnarray}

 \subsection{Input parameter scheme}
  In the input parameter scheme $\{G_F, M_Z, M_W\}$, the parameters in the Feynman rules 
  of the Warsaw basis 
 are given by, 
 \begin{eqnarray}
  v^2 &=& v^2_{\rm SM} \Big( 1 + \frac{1}{\sqrt{2} G_F} (2 C_{H\ell}^{(3)} - C_{\ell\ell})  \Big)\label{eqn:effective-v} \\
  g_1 &=& g_1^{\rm SM} \Big( 1 - \frac{1}{2\sqrt{2} G_F} (2 C_{H\ell}^{(3)} - C_{\ell\ell})
          - \frac{\delta m_Z^2}{2(M_Z^2-M_W^2)} \Big)\label{eqn:effective-g1} \\
  g_2 &=& g_2^{\rm SM} \Big( 1 - \frac{1}{2\sqrt{2} G_F} (2 C_{H\ell}^{(3)} - C_{\ell\ell})   \Big)\label{eqn:effective-g2}
 \end{eqnarray}
where, 
\begin{eqnarray}
 v^2_{\rm SM} &=& \frac{1}{\sqrt{2} G_F} \\
 g_1^{\rm SM} &=& 2 (2)^{(1/4)} \sqrt{G_F(M_Z^2-M_W^2)} \\
 g_2^{\rm SM} &=& 2 (2)^{(1/4)} M_W \sqrt{G_F}
\end{eqnarray}
and, corresponding changes in $s_W$ and $e$ should 
also be taken into account~\cite{Brivio:2017bnu}.
These values have to be used in the SM Feynman rules. In the Feynman
rules proportional to the Wilson  
coefficients, the parameters should be simply replaced by their SM
definitions. For SILH basis, one  
needs to replace $C_{H\ell}^{(3)} \to K'_{H\ell}/v^2~$ and
$C_{\ell\ell} \to K_{\ell\ell}/v^2$ in  
Eqs.~\ref{eqn:effective-v}-\ref{eqn:effective-g2}.

\section{Production and decay signal strengths}\label{appen:signal-strengths}

The signal strengths for production channels at $\sqrt{{\hat s}}=13~{\rm TeV}$ are 
given by, 
\begin{eqnarray}
\mu_{\rm VBF}  &=& 1.0 -  0.89\,c_{ZZ} - 2.5\,c_{Z\square} , \label{eqn:xsvbf} \\
\mu_{WH} &=& 1.0 + 4.6\,c_{ZZ} + 10.0\, c_{Z\square} , \label{eqn:xswh}\\
\mu_{ZH}  &=& 1.0 + 3.5\,c_{ZZ} + 8.3\,c_{Z\square}. \label{eqn:xszh}
\end{eqnarray}
The ggF and $ttH$ production channels do not depend on these parameters at the LO. 
The signal strengths for the decays are given by,
\begin{eqnarray}
\mu^{\gamma\gamma} &=& 1.0 +0.97 ~c_{ZZ} + 1.98 ~c_{Z\square}\label{eqn:aa}, \\
\mu^{2\ell2\nu} &=& 1.0 +0.044 ~c_{ZZ} + 0.52 ~c_{Z\square}\label{eqn:2l2nu}.
\end{eqnarray}
All the above expressions are taken from Refs.~\cite{Falkowski:2015fla, DiVita:2017eyz}. 
The partial decay widths for the $H \to 4\ell$ decay at the inclusive and differential levels 
are calculated using the {\tt Hto4l} code. The signal strengths in this channel are given by,  
%
%
\begin{eqnarray}
 \mu^{4\ell, \rm incl.} &=& 1.0 -0.19 ~c_{ZZ} + 0.21 ~c_{Z\square}\label{eqn:4l-incl}  \\
 \mu^{4\ell, \phi(1)} &=& 1.0 -0.12 ~c_{ZZ} + 0.31 ~c_{Z\square}\label{eqn:4l-bin1}  \\
 \mu^{4\ell, \phi(2)} &=& 1.0 -0.24 ~c_{ZZ} + 0.11 ~c_{Z\square}\label{eqn:4l-bin2}  \\
 \mu^{4\ell, \phi(3)} &=& 1.0 -0.12 ~c_{ZZ} + 0.32 ~c_{Z\square}\label{eqn:4l-bin3}, 
\end{eqnarray}
where $\phi(1), \phi(2)$ and $\phi(3)$ refer to the first ([0,90]), second ([90,270]) and third ([270,360]) bins 
of the $\phi$ distribution, respectively.

\bibliographystyle{JHEP.bst}
\bibliography{biblio}

\end{document}